\documentclass[aps,pra,showpacs,twoside,twocolumn,10pt]{revtex4-1}
\usepackage[colorlinks=true, citecolor=red, urlcolor=blue ]{hyperref}
\usepackage{xcolor}
\usepackage{epsfig,newlfont,amssymb,amsfonts,amsmath,bm,subfigure,palatino,mathtools,amsthm,braket,times,soul,enumitem,color}
\usepackage[normalem]{ulem}
\newcommand{\stkout}[1]{\ifmmode\text{\sout{\ensuremath{#1}}}\else\sout{#1}\fi}

\usepackage[utf8]{inputenc}
\usepackage{array}
\usepackage{graphics}

\newcommand{\ketbra}[2]{|#1\rangle \langle #2|}
\begin{document}

\title{Transient effects in quantum refrigerators with finite environments}
\author{Aparajita Bhattacharyya$^1$, Ahana Ghoshal$^{1,2}$, and Ujjwal Sen$^1$}
\affiliation{
$^1$Harish-Chandra Research Institute, A CI of Homi Bhabha National Institute, Chhatnag Road, Jhunsi, Prayagraj 211 019, India\\
$^2$Naturwissenschaftlich-Technische Fakult\"{a}t, Universit\"{a}t Siegen, Walter-Flex-Stra\ss e 3, 57068 Siegen, Germany}

\begin{abstract}
  We explore a small quantum refrigerator consisting of three qubits, each of which is kept in contact with an environment. We consider two settings: one is when there is necessarily transient cooling and the other is when both steady-state and transient coolings prevail. 
Our primary focus, however, is on the transient cooling phenomena. We show that in the transient regime, the temperature of the cold qubit can decrease further compared to the case where all qubits are connected to Markovian environments, by replacing the environment attached to the cold qubit with a finite-size spin environment, modeled by a few quantum spins interacting with the cold qubit.
 We also consider refrigeration with more than one finite-size spin environments 
 of the three-qubit refrigerating device.
  As expected, a steady temperature is reached only if there are at least two Markovian baths, regardless of whether the cold qubit is attached to an environment. However, distinct envelopes of temperature oscillations are observed in all cases. 
  We investigate the effects of the finite-size environments on the cooling of the cold qubit when one, two, or all Markovian baths are replaced by finite-size environments. Additionally, we examine this effect in two- and single-qubit self-sustained devices connected to one or more finite-size environments. 
  In deriving the dynamical equations for the qubits connected to finite-size environments, we made no assumptions about the Markovian nature of the environment. As a result, these finite-size environments inherently exhibit information backflow from the environment to the system, a hallmark of non-Markovianity. Hence, we propose a witness to detect non-Markovianity in  such systems. Finally, the cooling processes are studied  in presence of Markovian noise, and we analyse the response on the refrigeration of the noise strength. In particular, we find the noise strength until which refrigeration remains possible. 
\end{abstract}

\maketitle
\section{Introduction}
The onset and flourishing of quantum thermodynamics~\cite{Allahverdyan,Gemmer,Kosloff,Brand,Gardas,Gelbwaser,Misra,Millen,Vinjanampathy,Goold,Benenti,Binder,Deffner}
has sustained the arena of quantum thermal devices, which in turn has gained  utmost importance in the advancement of quantum technologies and in particular, for the miniaturization of quantum circuits. In the last few decades, the achievements in designing quantum heat engines and quantum refrigerators~\cite{Palao,Feldmann,Popescu,Levy1,Levy,Uzdin,Clivaz,Mitchison,Scarani}, quantum diodes~\cite{Yuan}, quantum thermal transistors~\cite{Joulain,Zhang,Su,Mandarino}, quantum batteries~\cite{Alicki_Fannes,Campaioli,Dutta} has built the backbone of quantum technologies, which aims to enhance the efficiencies of quantum devices over their classical counterparts~\cite{Kosloff2,Kosloff1,Wehner_new,Kurizki_new,Chen_Liu}. The confluence of quantum thermodynamics with many-body physics~\cite{Dorner,Mehboudi,Reimann,Eisert,Gogolin,Skelt}, quantum information theory~\cite{Gour,Vinjanampathy,Goold}, statistical and solid-state physics~\cite{Fazio,Rigol} have triggered the implementations of quantum devices in experiments using superconducting qubits~\cite{Pekola2,Aslan,Jordan}, mesoscopic substrates~\cite{Pekola}, ionic systems~\cite{ionic1,ionic2}, nuclear magnetic resonance~\cite{n_m_resonance}, etc. The performance of such devices is evidently controlled by thermal baths or  environments connected to the machinery setup, and the dynamics of the device components are driven by the open quantum evolutions~\cite{Petruccione,Alicki,Rivas,Lidar}. The thermal environments, that influence the efficiency of performance of the devices, can either be Markovian or non-Markovian depending on the validity of Born-Markov and secular approximations~\cite{Petruccione,Alicki,Rivas,Lidar}. In general, the efficiencies of quantum thermal devices immersed in Markovian~\cite{Petruccione} baths are computable and the dynamics of the machinery components 
can be efficiently handled.
Most of the thermal environments, however, reside in a non-Markovian family and therefore makes the realistic situations different from the ideal Markovian dynamics. There exists a significant body of work on quantum thermal machines operating under more than one thermal environments, which are either all Markovian~\cite{Palao,Levy1,Uzdin,Mitchison,Joulain,Su,Mandarino,Dutta} or all non-Markovian~\cite{Dutta,non-Markov,Kurizki,Uzdin1,Kato,Chen,Ostmann,Arpan,Shirai,Raja,Chakraborty,Carrega,Koyanagi,Filippis,Krzysztof}.  


The miniaturization of technologies has acquired a considerable momentum with the introduction of the concept of \textit{quantum absorption refrigerators} by Linden \textit{et al.} in~\cite{Popescu}. 
The devices consist of a small number of qubits~\cite{Popescu,Popescu2,Popescu3,Popescu4,Brask,Palao2,Palao3,Silva,Fazio2,Naseem,Woods,Sreetama,Ghoshal2,Chiara,Bhandari,Tanoy1,Tanoy3,Okane,Damas} and/or qudits~\cite{Popescu,Segal,Man,Wang,Tanoy2,Cao} which are driven by local Markovian baths attached to the respective subsystems and a local cooling of one of the qubits, referred to as the \textit{cold qubit}, can be attained. 
The absorption refrigerators are self-contained refrigerators, that usually operate in the absorption region, where no external energy is required for the cooling process~\cite{Popescu,Popescu2}. The dynamics of the device components is regulated to transfer thermal energy from a cold to a hot environment with the aid of a third thermal environment, known as the work reservoir, both in the steady and in the transient regimes, in order to decrease the cold qubit's temperature with respect to its initial temperature. To put it in another way, maintaining the state of the cold qubit in the currently accessible ground state allows the cooling of the qubit which is achieved by lowering the system's local temperature. 

Along with theoretical advancements, the implementation of quantum absorption refrigerators in quantum few-level systems have been devised by employing quantum dots~\cite{Giovannetti,Monsel}, atom-cavity systems~\cite{Plenio,Potts} and circuit QED architectures~\cite{Brask2}. Recently, three trapped ions~\cite{Scarani2} have also been used to construct a quantum absorption refrigerator. These refrigerators are anticipated to be helpful in instances where in-situ, on-demand cooling of systems as small as a qubit may be necessary without the need of external energy transfer and faster than the qubit's equilibriation time with a thermal environment.

The aforementioned works on quantum absorption refrigerators are mostly investigated in Markovian baths. Naturally, the efficiency of performance of the devices may be altered when connected with non-Markovian environments. In practical situations, most of the environments exhibit non-Markovian behavior. In order to belong to the Markovian family, the thermal environments must be infinitely large and have a continuous energy spectrum~\cite{Petruccione}. The bosonic environment, consisting of an infinite number of harmonic oscillators, within certain constraints, behaves as a Markovian one. Most of the common environments, such as spin-baths~\cite{Misra_Pati,Prokofev,Fisher_Breuer,Chitra,Majumdar}, are not Markovian. A few non-Markovian environments have Markovian limits, while for others, such as the spin star model, such a limit is evasive~\cite{Breuer1}. As Markovian nature of a thermal environment is, in general, far from the realistic scenario, it is important to study the effect of non-Markovianity on the refrigeration process. Sometimes, the situation may be more complex, in that while some of the thermal environments connected locally to the device components are Markovian, and the rest are not so. For such mixed local environments, the sub-systems of the machinery setup evolve under a combination of local Markovian and non-Markovian dynamics~\cite{Ghoshal}. Such situations can arise, e.g. while considering hybrid systems like atom-photon arrangements.
In this paper, we consider a few-qubit cooling process where each qubit is connected with a local reservoir, and look at the effect of substituting some or all Markovian reservoirs by finite-size spin environments. 
The Markovian baths are considered to be bosonic in nature, interacting with the appropriate qubits via Markovian qubit-bath interactions. When the Markovian baths are replaced by finite-size spin environments,  
the model  
demonstrates better transient cooling than the complete Markovian scenario, where we refer to the latter as  the ``ideal" case.
Note that, due to the small size of the finite-size environments, they cannot be considered as thermodynamic baths, which typically consist of an infinite number of particles. Instead, together with the refrigerator qubit, they form spin-star systems with a limited number of qubits, making them finite in size. Therefore, we do not refer to the environments as non-Markovian baths. However, as we did not make any assumptions about the Markovian nature of the environment when deriving the dynamical equations for the qubits attached to the spin environments, these finite-size environments inherently lead to information backflow from the environment to the system, which is a characteristic of non-Markovianity. In this work, non-Markovianity refers specifically to this information backflow caused by the finite size of the environments. There are some non-Markovian models for infinite baths~\cite{nm1,nm2,nm3}, but they are not considered in this study. Additionally, it is important to note that because the environments are finite in size and, consequently, have finite energy, they are not suitable for the steady-state refrigerator model, because they cannot supply energy for continuous cooling. They are, however, a good fit for achieving refrigeration during transient times. Thus, in this work, we primarily focus on the cooling of the cold qubit during the transient time. Transient refrigeration has already been explored previously, though those studies were limited to Markovian dynamics~\cite{Sreetama}. Along with three-qubit cooling processes, with one or more 
finite-size environments,  
we also, for comparison and completeness, consider
single- and two-qubit self-sustained thermal devices kept in contact with one or more spin-environments, which in certain situations also exhibit refrigeration.
Subsequently, we propose a 
witness
of non-Markovianity in these devices.
Finally, since noise permeates all practical implementations of quantum machines, the three-qubit cooling process is analysed in presence of several Markovian noise models. 

The remainder of the paper is arranged as follows. The relevant information necessary to formulate the problem is discussed in Sec.~\ref{Sec:2}. This includes identifying the system Hamiltonian, the initial state, and providing a formal definition of local temperature of the individual qubits.
In Section~\ref{non_mark}, we analyse 
the interaction of the system with the finite-size spin environments
with a detailed description of the system operators. 
In Sec.~\ref{m_nm}, the effects of mixed Markovian and non-Markovian environments 
on refrigeration are discussed.
In Sec.~\ref{Sec:4}, we provide 
a  witness of non-Markovianity, and compare the same
with the well-known RHP witness.
Sec.~\ref{Sec:5} illustrates the operation of the cooling process in presence of noise. Finally, the concluding remarks are presented in Sec.~\ref{Sec:6}.

\begin{figure}
\includegraphics[width=8cm,height=5.5cm]{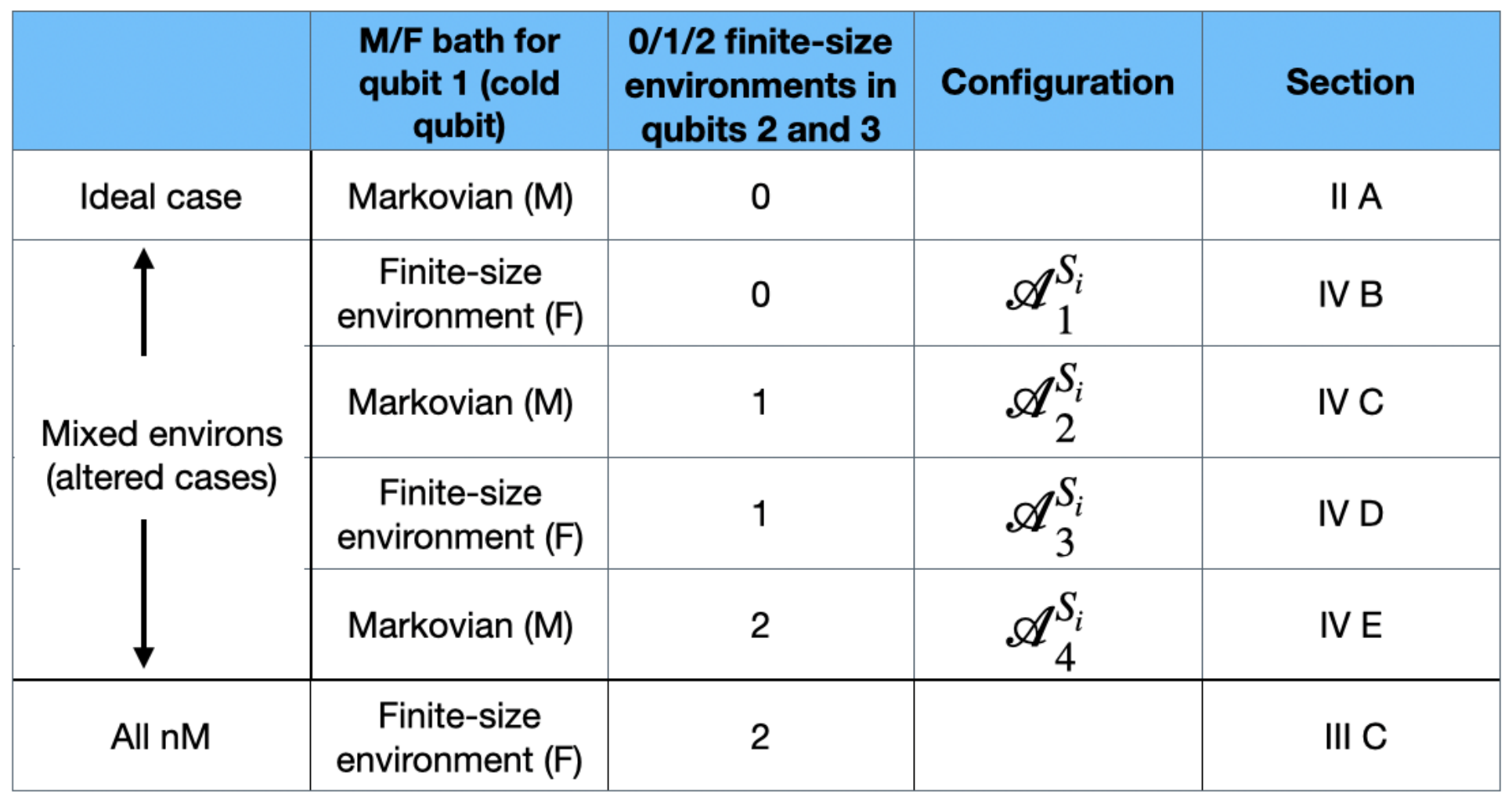}%

    \caption{We present here a summary of the different cases considered in the paper, and point to the respective sections where they are discussed. For each of the cases, we consider three setups, viz. $S_1$, $S_2$ and $S_3$. Similarly, the two-qubit and single-qubit cases are also considered. The abbreviations, $M$ and $F$ denote the Markovian and finite-size environments respectively.}
    \label{tabular}
\end{figure}

\section{Quantum absorption refrigerator}
\label{Sec:2}
We begin with a formal description of the quantum absorption refrigerator, which consists of three interacting qubits, each locally and separately connected to three thermal environments denoted as $B_1^X$, $B_2^X$, and $B_3^X$, where $X\in \{M,NM\}$. Here, ``$M$" and ``$NM$" represent Markovian and non-Markovian environments, respectively. The composite system-environment configuration is governed 
by the Hamiltonian,
\begin{equation}
    H=H_S+H_{B^X}+H_{SB^{X}}.
    \label{eq:Hamil}
\end{equation}
Here, $H_S$ stands for the sum of the local Hamiltonians of the three qubits of the refrigerator, while $H_{B^X}$ represents the sum of  the local Hamiltonians of the  three  environments. The term $H_{SB^X}$ denotes the interactions between the system qubits and the environments. 
The first qubit, often referred to as the ``cold qubit", is the target for cooling, whereas the second and third qubits perform the refrigeration~\cite{Popescu}. 

The Hamiltonian describing the three-qubit composite system can be expressed as $H_S = H_{loc} + H_{\text{int}}$, where $H_{loc}$ is given by $H_{loc} = \sum_{i=1}^3 H_{S_i}$. In this context, each $H_{S_i}$, where $i$ ranges from $1$ to $3$, represents the local Hamiltonian for each of the three individual qubits, and $H_{\text{int}}$ characterizes the interactions between them. The specific forms of $H_{S_i}$ and $H_{\text{int}}$ are given by
\begin{eqnarray}
    &&H_{S_i}=\frac{K}{2} E_i \sigma_i^z,\nonumber\\
    &&H_{\text{int}}=Kg(\ketbra{010}{101}+\ketbra{101}{010}).
    \label{eq:Hamil1}
\end{eqnarray}
Here, the states $\ket{1}$ and $\ket{0}$ correspond to the ground and excited states of the qubits, respectively. These states have energies, $-\frac{K E_i}{2}$ and $\frac{K E_i}{2}$, corresponding to the $i^{\text{th}}$ qubit, with $K$ being a constant parameter having the unit of energy.
The dimensionless parameters, $\{E_i\}$, represent the energy levels of the 
same, and $g$ is the dimensionless interaction strength. Therefore, the energy parameters and the interaction strength throughout the manuscript are expressed in units of $K$, which normalizes the energy scale. $\sigma_i^z$ stands for the $z$-component of the Pauli matrices $\vec{\sigma_i}(\sigma_i^x,\sigma_i^y,\sigma_i^z)$.
The individual qubits are connected with local heat environments of temperatures $\tau_1$, $\tau_2$ and $\tau_3$ respectively, where $\tau_1 \le \tau_2 \le \tau_3$. 
$\tau_2$ is initially set as the room temperature. Note that,  $\tau_1$, $\tau_2$, and $\tau_3$ are dimensionless temperatures, whereas the actual temperatures, $\tilde{\tau}_1$, $\tilde{\tau}_2$, and $\tilde{\tau}_3$, are related to these dimensionless temperatures as $\tau_1=\frac{k_B\tilde{\tau}_1}{K}$, $\tau_2=\frac{k_B\tilde{\tau}_2}{K}$, and $\tau_3=\frac{k_B\tilde{\tau}_3}{K}$, with $k_B$ representing the Boltzmann constant.

In this work, we aim to construct a self-contained refrigerator that operates autonomously, independent of any external energy source. To achieve this, it is essential to understand algorithmic cooling within this context~\cite{ac1, ac2}. Algorithmic cooling is based on three interconnected processes. First, there is a manipulation of the local entropy of certain computational bits, which causes some to cool while others become significantly hotter than their surroundings. Next, through controlled interactions, polarization is transferred from the reset bits to the computational bits, allowing the hotter computational bits to adiabatically lose their entropy to the reset bits. Finally, the cooler computational bits remain isolated, while the reset bits quickly revert to their initial states, releasing their accumulated entropy into the environment. In a two-qubit cooling process, the cooling of the first qubit is achieved by increasing its ground state population. However, the two-qubit model of algorithmic cooling is not self-sufficient and requires an external source of energy, typically provided by a unitary operator. For example, in NMR experiments, this external energy can be supplied via a sequence of magnetic field pulses~\cite{ ac2,nmr}.
In the seminal work by Popescu et al.~\cite{Popescu}, they proposed a model in which the external energy is supplied by a bath at a higher temperature, acting as a source of free energy. In this setup, the qubit connected to the higher-temperature bath serves as the ``engine" of the cooling process. The process involves qubits 2 and 3 acting as the spiral and engine, respectively, while qubit 1 is the object being cooled. To make the cooling process self-contained, the condition $E_1+E_3=E_2$ must be satisfied. Under this condition, the energy levels $\ket{010}$ and $\ket{101}$, become degenerate, enabling the transition between $\ket{010} \leftrightarrow \ket{101}$, which allows for a self-sustained cool ing process without the need for an external energy source. This framework for algorithmic cooling applies broadly to various environments, including Markovian and non-Markovian baths, and hence is also valid for finite-size spin environments. Therefore, we use this model to study refrigeration with finite-size spin environments in our work.  
Additionally, we also set the first qubit to remain at room temperature initially, i.e.,  $\tau_1=\tau_2$.

At the initial time $t=0$, let us consider a scenario in which the three qubits are  independently in thermal equilibrium with each of the three reservoirs that are locally connected to them. Here, $t$ denotes the dimensionless time with the actual time $\tilde{t}$, defined as $t=\frac{K \tilde{t}}{\hbar}$. 
So, the initial state of the combined three-qubit system is 
given by
\begin{equation}
    \rho_0=\rho_0^1 \otimes \rho_0^2 \otimes \rho_0^3,
\end{equation}
where $\rho_0^i=Z_i^{-1}\exp(-\beta_i E_i \sigma_i^z/2)$, with $Z_i$ being the partition function for the $i^{th}$ qubit given by $Z_i=\text{Tr}[\exp(-\beta_i E_i \sigma_i^z/2)]$, and $\beta_i$ is the corresponding dimensionless inverse temperature expressed as $\beta_i=1/\tau_i$.
After  activation of the interaction between the qubits for $t>0$, the system undergoes open quantum dynamics, which is described by the quantum master equation~\cite{Petruccione, Alicki, Rivas, Lidar} given as
\begin{equation}
\label{QME}
    \frac{\partial \rho_s(t)}{\partial t} =\mathcal{L}(\rho_s(t))= -\frac{i}{K}[H_S,\rho_s(t)] +\sum_{i=1}^3 \frac{\hbar}{K}\mathcal{D}_i(\rho_s(t)),
\end{equation}
where $\rho_s(t)$ is the reduced state of the composite three-qubit system at time $t$ and $\mathcal{D}_i(\rho_s(t))$ represents the dissipative term accounting for the decoherence effects originating from the $i^{\text{th}}$ environment. The specific form of $\mathcal{D}_i(\rho_s(t))$ depends entirely on the type of the $i^{\text{th}}$ environment connected to the systems, and it may exhibit distinct characteristics for Markovian and non-Markovian environments. 
In the literature, a quantum absorption refrigerator is usually studied by using Markovian environments, i.e., with three Markovian environments, $B_1^M$, $B_2^M$, and $B_3^M$.
Initially, the density matrices of the three reduced subsystems are diagonal in the eigenbasis of $H_{S_i}$. As Markovian environments do not induce quantum coherence, the local subsystems $\rho_i(t) = \text{Tr}_{j,k}\big(\rho_s(t)\big)$, where $j,k \ne i$ and $i,j,k \in {1,2,3}$, also remain diagonal in the same basis, as time progresses. This property allows us to define the local temperatures for the individual qubits as follows. Let 
\begin{equation}
    \rho_1(t)=r_1(t) \ketbra{0}{0} +\big[1-r_1(t)\big]\ketbra{1}{1},
\end{equation}
be the time-evolved reduced state of the cold qubit, where $r_1(t)$ is the population in the excited state at time $t$. The local (time-dependent) temperature, \(T_1(t)\) of the cold environment can then be given by $r_1(t)=\frac{1}{Z_1(t)}\exp (-E_1/2 T_1(t))$, with $Z_1(t)=\text{Tr}\big[\exp(- E_1 \sigma_1^z/2T_1(t))\big]$. $T_1(t)$ is 
dimensionless, and
is related to the actual temperature $\tilde{T}_1(t)$  by
\begin{equation}
\label{temp}
    T_1(t)=\frac{k_B \tilde{T}_1(t)}{K}=E_1\Big[ \ln\Big( \frac{1-r_1(t)}{r_1(t)}\Big) \Big]^{-1}.
\end{equation}
The local temperatures of the remaining two qubits can also be defined in a similar fashion. Thus, the definition of local temperatures for any qubit is based on the populations of ground and excited states of the system. A decrease in local temperature of any qubit is here manifested as an increase in the ground state population of that qubit. We will use the same  definition of temperature even while using non-Markovian environs. This suits our purpose as the non-Markovian environments used in this paper also do not induce quantum coherence -- in the (energy) eigenbasis of \(H_{S_i}\) -- in the local qubits. However, the definition of the local temperature presumes that \(r_1(t)\) is no higher than \(1-r_1(t)\).
In all further discussions, temperature and time indicate the corresponding dimensionless temperature and time defined above.

For proper refrigeration to occur at time $t$, it is necessary for the local temperature of the cold qubit to be significantly lower than its initial temperature, denoted as $T_1(t) < \tau_1$.
If the temperature of the cold qubit in the steady state, denoted as $T_1^S$,  is lower than $\tau_1$, we refer to this as the achievement of a steady state cooling (SSC). Also, in the transient regime, cooling can be achieved within the time frames shorter than that required to reach a steady state, resulting in temperatures considerably lower than $\tau_1$. This type of cooling is termed as transient cooling (TC). 
Sometimes TC can be obtained without the occurrence of SSC~\cite{Sreetama}.

\subsection{Refrigeration in presence of Markovian environments}
\label{Sec:3}
A local cooling of qubit $1$ can be obtained by the three-qubit three-environment setup with the environment configuration $\big\{B_1^M,B_2^M,B_3^M\big\}$. This configuration represents the situation where all three environments are Markovian in nature. Consequently, the Hamiltonian describing the entire setup is now determined by Eq. (\ref{eq:Hamil}), where $X$ is replaced here by $M$, to indicate that the environments are now Markovian.
We consider the Markovian baths as bosonic reservoirs composed of an infinite number of harmonic oscillators. The Hamiltonian describing these baths can be expressed as  $H_{B^M}=\sum_{i=1}^3 H_{B_i^M}$, where
\begin{equation}
    H_{B_i^M}=
    \int_{0}^{\Omega}\hbar\tilde{\omega}\eta_{\omega}^{i\dagger}\eta_{\omega}^id\omega.
    \label{eq:Mar_bath}
\end{equation}
Here $\tilde{\omega}$ represents an arbitrary constant with the dimensions of frequency, while $\Omega$ denotes the cutoff frequency, which is assumed to be the same for all the baths. This cutoff frequency is set to be sufficiently high so that the memory time of the bath $\sim \Omega^{-1}$, is negligibly small. This choice of parameters allows us to apply the Markovian approximations~\cite{Petruccione,Alicki,Rivas,Lidar}. The operators $\eta^{i\dagger}_{\omega}(\eta^i_{\omega})$, having the unit of $\frac{1}{\sqrt{\omega}}$, represents the bosonic creation (annihilation) operators corresponding to the mode $\omega$ of the $i^{\text{th}}$ bath. The Hamiltonian describing the interaction between the system and the Markovian environments is given by $H_{SB^M}=\sum_{i=1}^3 H_{SB_i^M}$, where
\begin{equation}
\label{mark_int}
    H_{SB_i^M}=\int_0^{\Omega} \hbar \sqrt{\tilde{\omega}}d\omega h_i(\omega)\sigma_i^x\otimes (\eta_{\omega}^{i\dagger}+\eta^i_{\omega})
\end{equation}
describes the interaction between the $i^{\text{th}}$ system and the $i^{\text{th}}$ bosonic environment. Here $h_i(\omega)$ is a dimensionless function of $\omega$, that tunes the coupling between the $i^{\text{th}}$ qubit and $B_i^M$. For harmonic oscillator environments, $\tilde{\omega} h_i^{2}(\omega)=J_i(\omega)$, where $J_i(\omega)$ represents the spectral density function of $B_i^M$. In this paper, we have taken $J_i(\omega)$ to be Ohmic spectral density function in the form $J_i(\omega)= \alpha_i \omega \exp(-\omega/\Omega)$. Here $\alpha_i$'s stand for the dimensionless qubit-environment interaction strengths.
The dynamical equation for the system is described by the GKSL master equation, as
presented in Eq.~(\ref{QME}), with the dissipative terms,
\begin{equation}
\label{Lindblad}
    \mathcal{D}_i(\rho_s(t)) = \sum_{\omega^{\prime}} \gamma_i(\omega^{\prime}) \Big[L_i^{\omega^{\prime}} \rho_s(t) L^{\omega^{\prime\dagger}}_{i} -\frac{1}{2} \{ L_i^{\omega^{\prime\dagger}} L^{\omega^{\prime}}_{i}, \rho_s(t) \} \Big],
\end{equation}
with $i=1,2,3$. $\gamma_i(\omega^{\prime})$ is the decay constant with units of time$^{-1}$, and ${L^{\omega^{\prime}}_i }$ represents the Lindblad or jump operators associated with the possible transition frequencies $\omega^{\prime}$ of the system. 
Note that all the studies presented in this work regarding Lindblad-type dynamics are based on a global Lindblad master equation. In particular, the jump operators in the master equation are derived by decomposing the interaction Hamiltonian of the system and environment in the eigenbasis of the total three-qubit system Hamiltonian, along with the interactions between these qubits.
In order to validate the Born-Markov approximations, we operate within the framework of the weak coupling limit, i.e., $\max\{\tilde{\gamma}_i(\omega^{\prime})\} \ll \min \{E_i,g\}$, where $\tilde{\gamma}_i(\omega^{\prime})=\frac{\hbar \gamma_i(\omega^{\prime})}{K}$. If $\epsilon$ and $\epsilon^{\prime}$ are any two eigenvalues of $H_S$, then the transition frequency associated with the transition between these two energy eigenstates is defined as, $\omega^{\prime}=\frac{1}{\hbar}(\epsilon^{\prime}-\epsilon$), and for the construction of the corresponding Lindblad operators see~\cite{Petruccione,Alicki,Rivas,Lidar}. For our case, the explicit expressions of the decay constants $\{\gamma_i(\omega^{\prime})\}$ and the Lindblad operators $\{ L^{\omega^{\prime}}_i \}$ are provided in Appendix~\ref{appen:1}. Here, we make the assumption that the environments are initially in their thermal equilibrium state.

\begin{figure*}
\includegraphics[width=5.9cm]{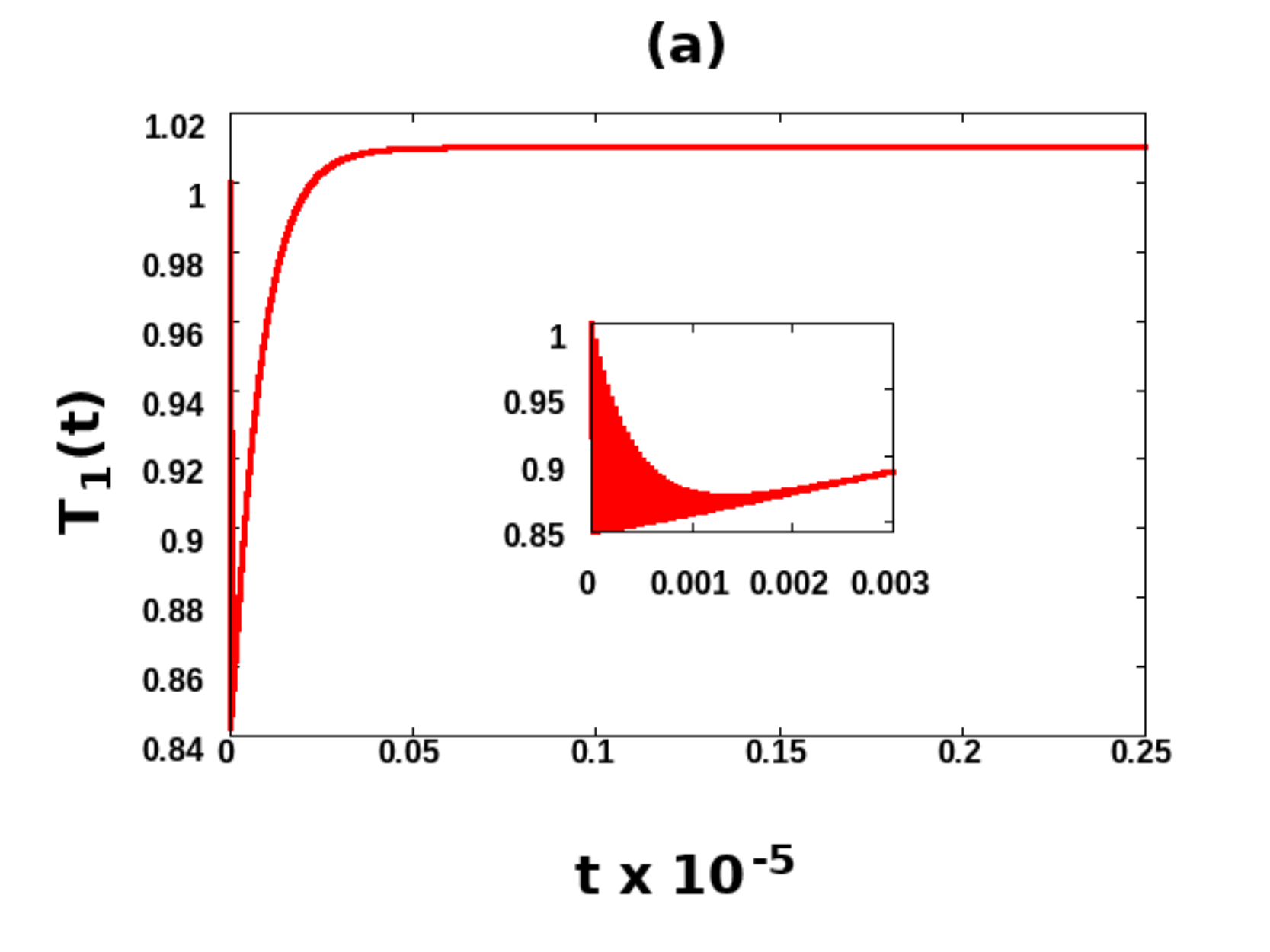}
\includegraphics[width=5.9cm]{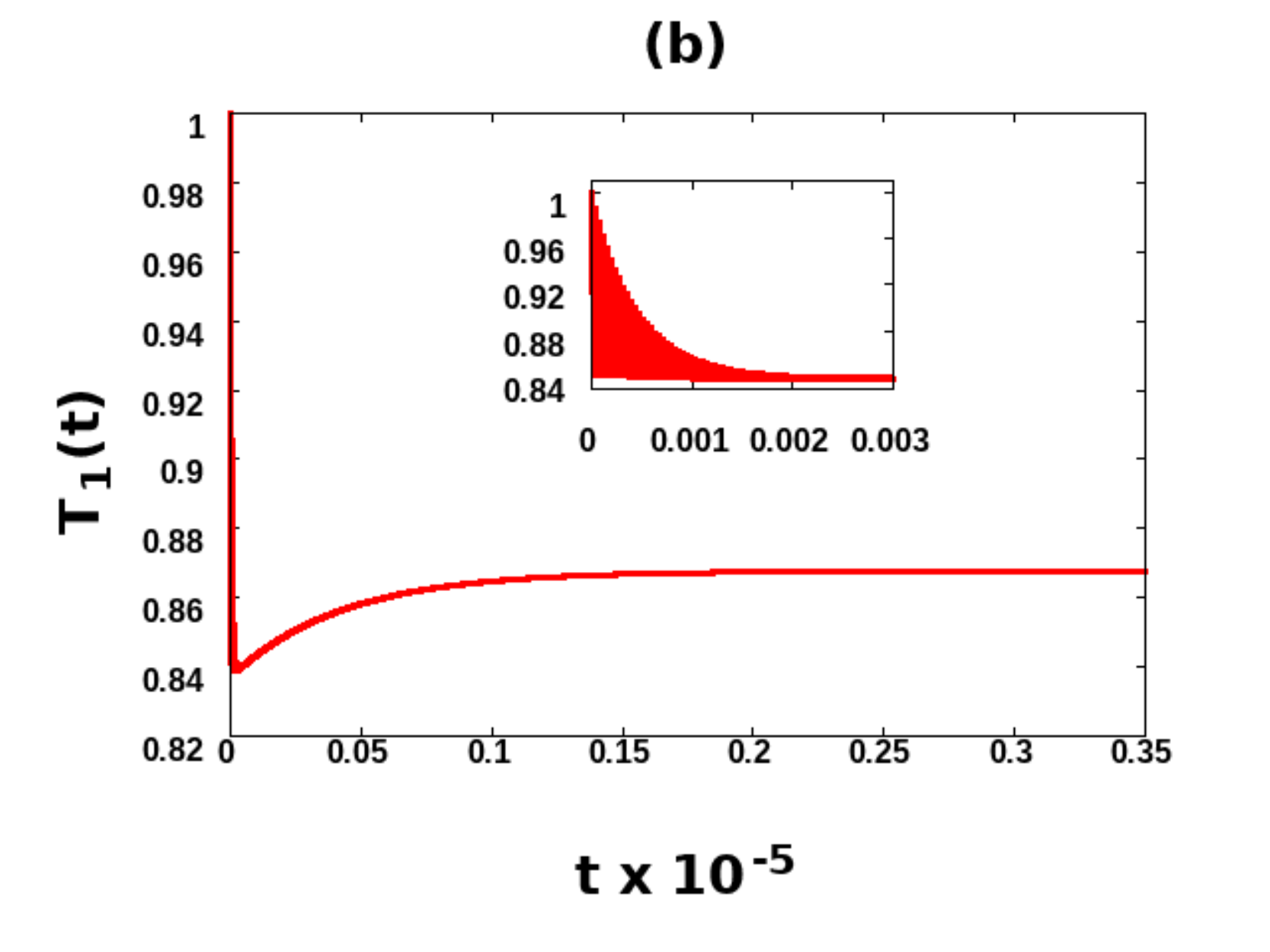}
\includegraphics[width=5.9cm]{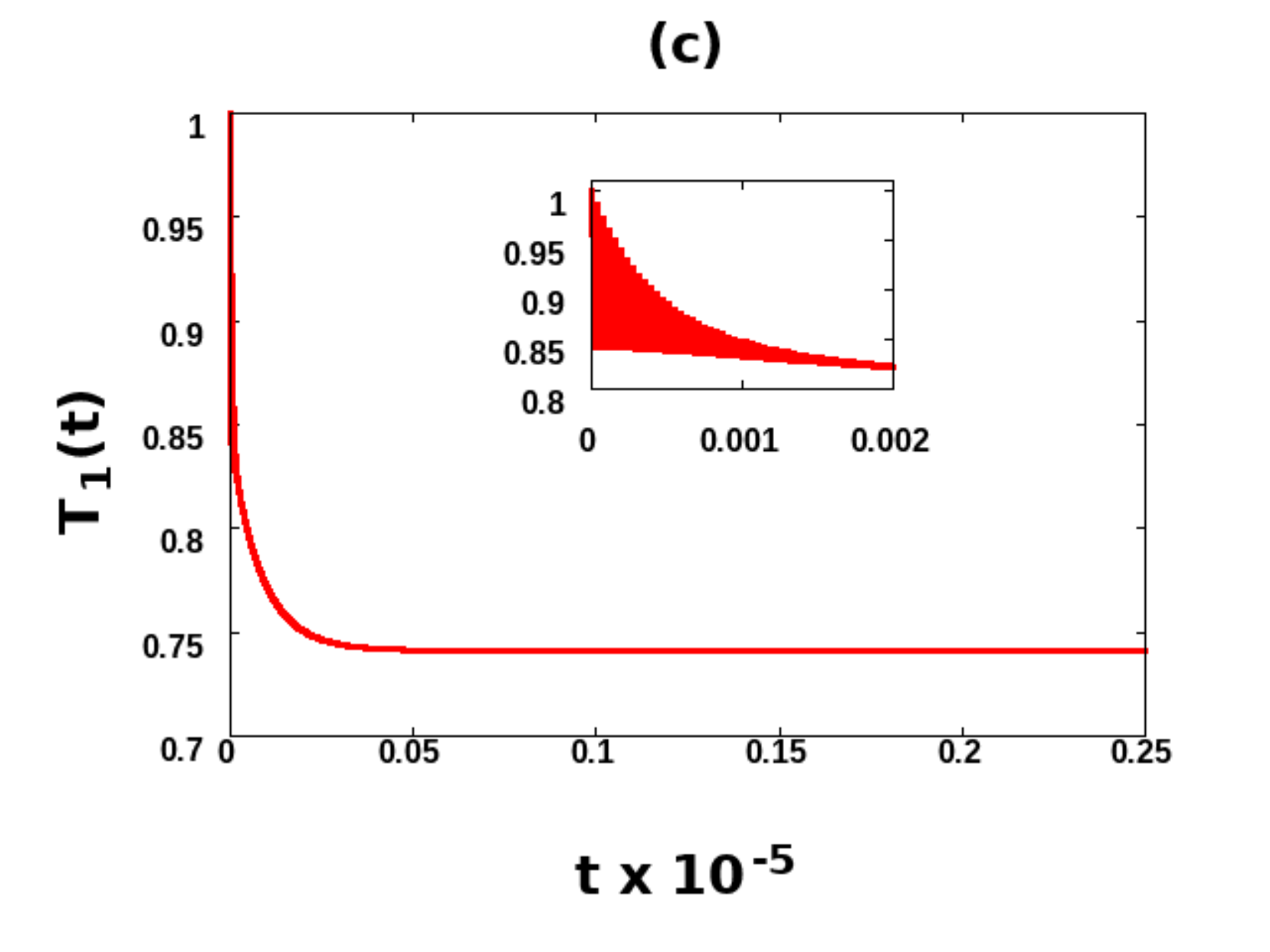}
\caption{Time dynamics of the cold qubit temperature of a three-qubit three-environment model of a quantum absorption refrigerator with the environment configuration $\big\{B_1^M,B_2^M,B_3^M\big\}$. Here we depict the dependence of $T_1(t)$ on $t$ in the three operating regimes: (a) $\mathbf{S_1}$: for $\alpha_1=10^{-3}$, $\alpha_2=10^{-4}$, $\alpha_3=10^{-2}$, (b) $\mathbf{S_2}$: for $\alpha_1=10^{-4}$, $\alpha_2=10^{-4}$, $\alpha_3=10^{-2}$, and (c) $\mathbf{S_3}$: for $\alpha_1=10^{-4}$, $\alpha_2=10^{-3}$ and $\alpha_3=10^{-2}$. The transient behavior near $t\times 10^{-5} \approx 0$ is shown in the insets of the main figures in the three panels. The system parameters are chosen to be $E_1=E_3=1.0$ and $g=0.8$. The temperatures of the thermal environments are taken as $\tau_1=\tau_2=1.0$ and $\tau_3=2.0$ and the cut off frequency of all the environments is taken to be $\Omega=10^3 \frac{K}{\hbar}$. All quantities plotted are dimensionless.}
\label{3mark}
\end{figure*}
It is already known that an ideal three-qubit three-environment quantum absorption refrigerator, configured as 
$\big\{B_1^M, B_2^M, B_3^M\big\}$, operates in three distinct regimes. These regimes depend on the qubit-environment interaction strength $\alpha_i$ for $i=1,2,3$, as discussed in~\cite{Sreetama}. Here, we provide an example for each of these three scenarios, while assuming strong coupling between the qubits, i.e., $g \approx E_i$.\\
\\
\textbf{$\mathbf{S_1}$: TC without SSC.} Fig.~\ref{3mark}-(a) shows the refrigeration of the cold qubit within a regime where transient cooling takes place, but steady-state cooling does not occur.
The minimum temperature during this transient phase is reached at approximately $\overline{T}_1^C \approx 0.84$. Subsequently, after a certain period, the temperature of the cold environment stabilizes at a temperature higher than its initial one, $\tau_1=1$, displaying a characteristic feature of \textit{steady state heating}. \\
\\
\textbf{$\mathbf{S_2}$: TC better than SSC.} From Fig.~\ref{3mark}-(b) we observe a scenario where both transient and steady-state cooling occur, but the transient cooling is better than the steady state one. In the transient regime, the behavior of the system qualitatively resembles that of the case $\mathbf{S_1}$ (see Fig~\ref{3mark}-(a)), reaching a minimum temperature at approximately $T_1^{\prime C}\approx 0.84$. Also, there is an additional feature of steady state cooling with the steady state temperature $T_1^{\prime S}\approx 0.88$, a feature entirely absent in the operating regime $\mathbf{S_1}$.
\\
\\
\textbf{$\mathbf{S_3}$: SSC better than TC.} In Fig~\ref{3mark}-(c) we depict an operating regime of the refrigerator in which both transient and steady-state cooling take place. However, in this case, steady-state cooling outperforms transient cooling, with the steady-state temperature reaching approximately $T_1^{\prime\prime S}\approx 0.75$.\\
\par
These three situations, denoted as $\mathbf{S_1}$, $\mathbf{S_2}$ and $\mathbf{S_3}$, and pertaining to the environment configuration $\big\{B_1^M,B_2^M,B_3^M\big\}$ will be referred as the ``ideal" setup in the subsequent discussions of this paper. In all three of these investigations, we identify specific parameter regimes, expressed as $\big\{\alpha_1,\alpha_2,\alpha_3\big\}$, for the operation of the refrigerators. For the convenience of notation, in the further discussions we will denote the specific regions as $\big\{\alpha_1^{S_i},\alpha_2^{S_i},\alpha_3^{S_i}\big\}$ for the corresponding situations $\mathbf{S_i}$ with $i=1,2$ and $3$. 
\section{What if all qubits are attached to finite-size spin environments?}
\label{non_mark}
\begin{figure*}
\centering
\includegraphics[width=6cm,height=4.2cm]{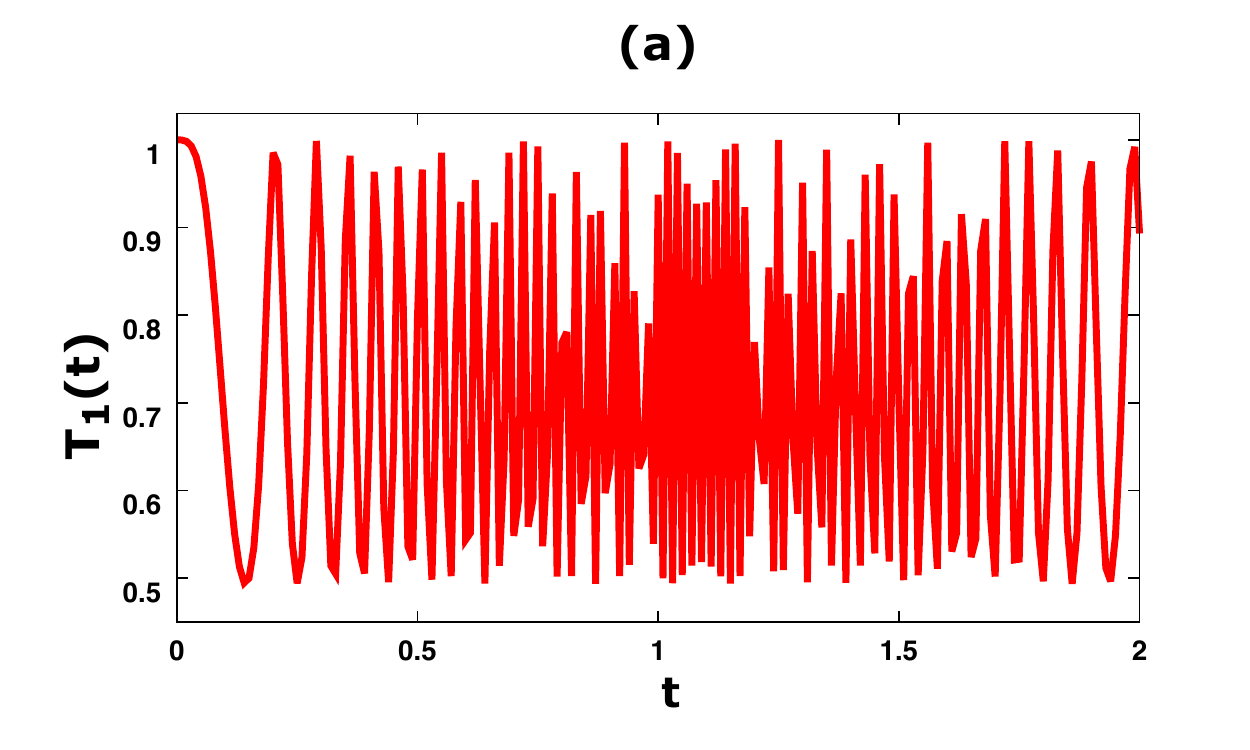}%
\includegraphics[width=5.8cm]{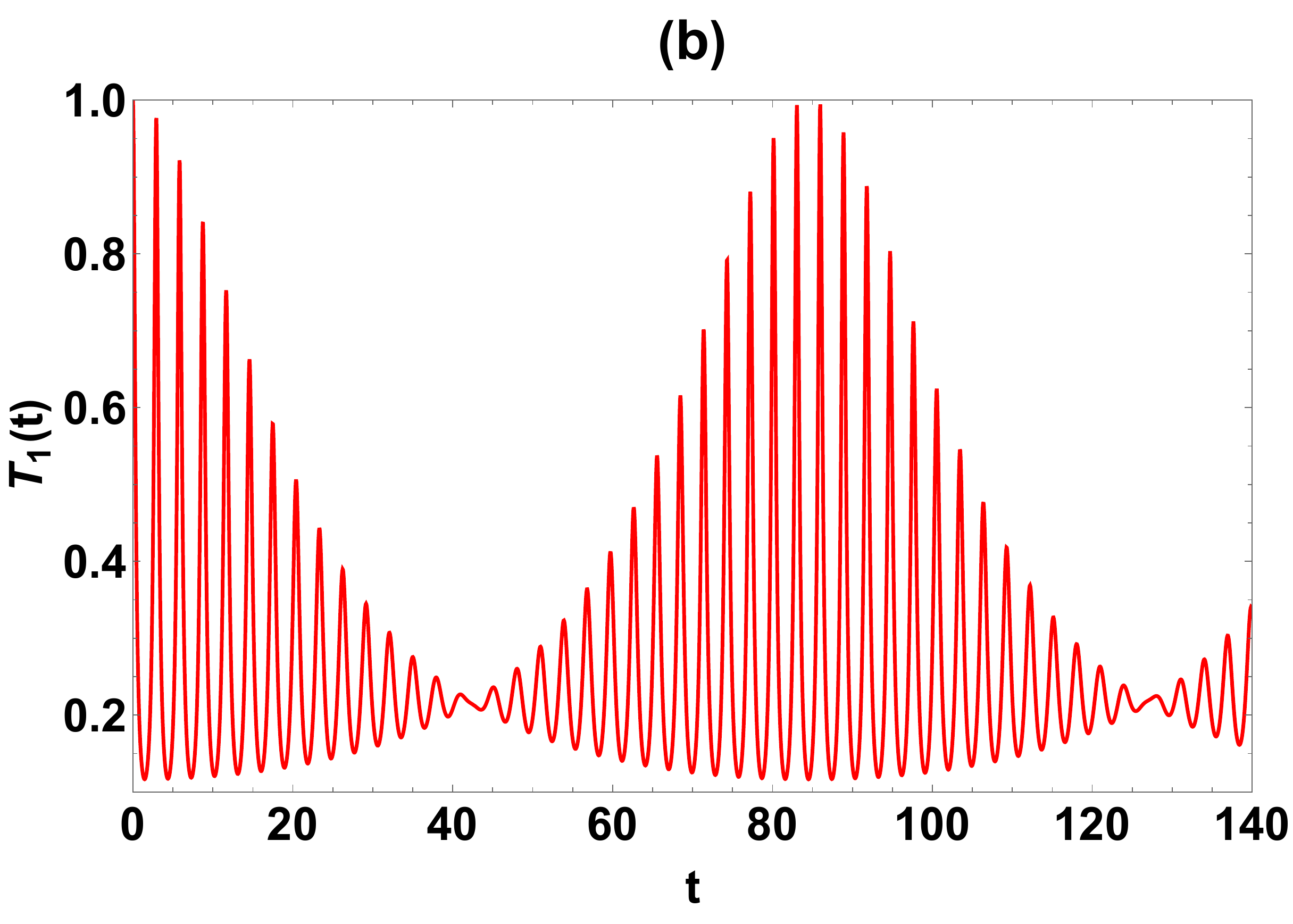}
\includegraphics[width=5.8cm]{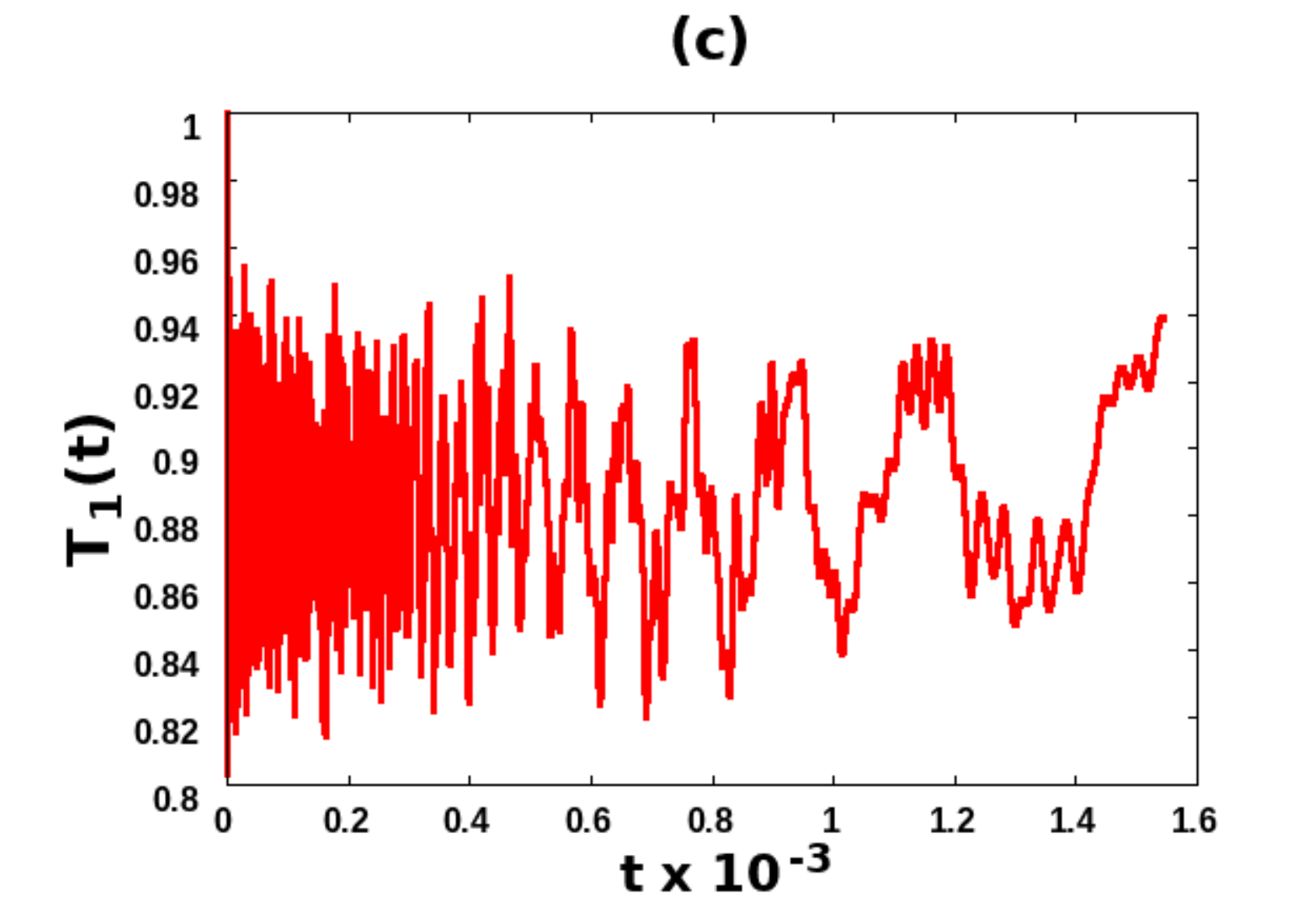}
\caption{Refrigeration of single-, two- and three-qubit quantum refrigerators in presence of non-Markovian environments. In panel (a), we depict the single-qubit scenario and plot the temperature of the qubit, $T_1(t)$, as a function of time, $t$. Here, $E_1=1/2$ and $N=2$.
In panel (b), we depict the cold qubit temperature $T_1(t)$ for $E_1=E_2=0.1$ and $N=2$. 
In panel (c), we present the same for a three-qubit three environment cooling process 
with the environment configuration $\big\{B_1^{NM},B_2^{NM},B_3^{NM}\big\}$. 
The system parameters and the temperatures of the environments are taken to be the same as in Fig.~\ref{3mark} 
and we set $N=2$. All  quantities plotted along the horizontal and the vertical axes are dimensionless. 
}
\label{1q_2q}
\end{figure*}
A Markovian scenario is highly specific and imposes strict constraints on the thermal environments, while the presence of non-Markovian environments 
in nature is more probable.
Therefore, it is necessary to investigate the impact of non-Markovian environments 
on refrigeration.
In this paper, we consider finite-size spin-environments, where we choose the system environment interaction to form a ``spin-star" model. 
As discussed earlier, these environments inherently exhibit non-Markovian behavior, as we make no assumptions about Markovianity when deriving the dynamical equations for the qubits connected to the spin environments. The spin-star model
consists of a total of $N+1$ spins, with $N$ of them arranged on the surface of a sphere at an equal distance from a central spin. The central spin represents the open system and is assumed to be described on a two-dimensional Hilbert space 
\(\mathcal{H}_A\), while the surrounding spins constitute the environment and are associated with a Hilbert space \(\mathcal{H}_B\), which
is an 
$N$-fold tensor product of two-dimensional spaces.
The Hamiltonian of the environments, as denoted by $H_{B^X}=\sum_{i=1}^3 H_{B_i^X}$ in Eq.~(\ref{eq:Hamil}), is now associated with non-Markovian spin environments. Therefore, we represent $X$ as $NM$, and the local Hamiltonian for the $i^{\text{th}}$ 
spin environment is given by
\begin{equation}
\label{nm_bath}
     H_{B_{i}^{NM}}=\hbar \nu_i J_{i}^+J_{i}^-,
\end{equation}
where
\begin{equation}
    J_i^{\pm}=\sum_{k=1}^{N} \sigma_{i}^{k\pm} \quad \text{and} \quad \sigma_i^{k\pm}=\frac{\sigma_{x_i}^k \pm i\sigma_{y_i}^k}{2},
    \label{eq:spin_bath}
\end{equation}
with $\nu_i$ being the frequency of $B_i^{NM}$.
The interaction between the central spin and its neighboring spins is modeled as the Heisenberg $XY$ interaction, as described in~\cite{Breuer1}, and is given by
\begin{equation}
\label{nm_int}
    H_{SB^{NM}}=\sum_{i=1}^3 H_{SB_i^{NM}}=2 K\alpha_0\sum_{i=1}^3 \Big( \sigma_i^+ J_i^- + \sigma_i^- J_i^+ \Big).
\end{equation}
Here, $\alpha_0$ stands for the dimensionless interaction strength, and for our entire analysis, we have chosen $\alpha_0=1/2$. It is important to note that the $N$ spins  located on the surface of the sphere do not interact with each other. The time dynamics of the reduced three-qubit system after tracing out the 
environments is controlled by the equation,
\begin{equation}
\label{nm_evol}
    \rho_s(t) = \text{Tr}_B \big[e^{-i\frac{\tilde{H}}{K}t} \rho(0) e^{i\frac{\tilde{H}}{K}t}\big],
\end{equation}
where $\rho(0)=\rho_0 \otimes \rho_{B_1} \otimes \rho_{B_2} \otimes \rho_{B_3}$ is the combined state of system-environment setup at time $t=0$, and $\tilde{H}=H_S+H_{SB^{NM}}$. Note that $H_{B^{NM}}$ does not contribute in the reduced dynamics of the system. Here also, the initial states of the 
environments are assumed to be in thermal equilibrium at their respective temperatures, i.e.,
$\rho_{B_i}=\frac{\exp(-\beta_i H_{B_i^{NM}})}{\text{Tr}\{\exp(-\beta_i H_{B_i^{NM}})\}}$.
For this particular choice of non-Markovian 
environments, the reduced subsystems of the qubits remain diagonal in the energy eigenbasis of $H_{S_i}$, at any time $t$. So, for this non-Markovian case also, we can define the local temperature of a qubit as in Eq.~(\ref{temp}). 


In the case of Markovian environments, it has been demonstrated that the smallest self-contained quantum absorption refrigerator is the three-qubit three-environment cooling process. It is not possible to design a cooling process with fewer than three qubits that can operate without any external control. This was initially established using a toy model and later confirmed for Markovian environments as well~\cite{Popescu}. However, for finite-size spin 
environments, the situation may differ. The presence of 
finite-size spin environments, which can lead to backflow effects, may enable the construction of a self-contained cooling process with fewer than three qubits. In the subsequent sections, we will explore the influence of finite-size spin 
reservoirs on the three-qubit three-environment cooling process. However, before delving into this scenario, we will first investigate the same for one and two-qubit quantum cooling processes. The objective is to ascertain whether it is possible to construct a self-contained cooling process with less than three qubits in the presence of finite-size spin 
environments. The two subsections below consider respectively single- and two-qubit cooling processes,  where all environments consist of a finite number of spins. 
For completeness, we also discuss the three-qubit cooling process with three finite-spin environments 
in Sec.~\ref{ekhane}.
\\

\subsection{Refrigeration with  single-qubit 
single-environ setup}
\label{1q}
Let us begin by considering a two-level system attached to a spin-environment, which is described by a spin-star model as detailed in Eq.~(\ref{nm_bath}). The qubit-environ 
interaction follows the form presented in Eq.~(\ref{nm_int}) for $i=1$. The Hamiltonian of the system is represented by $H_{S_1}$, defined in Eq.~(\ref{eq:Hamil}), and the initial combined 
system-environ state is taken as $\rho_0^1 \otimes \rho_{B_1}$,
a product of the thermal states of the qubit and 
environment, both of which are at the same temperature.
Following the time evolution of the system according to Eq.~(\ref{nm_evol}), we observe that the final state is also diagonal. We then proceed to calculate the temperature, $T_1(t)$, of the qubit  at time $t$. If we consider the environment to consist of only a single spin ($N=1$ in Eq.~(\ref{eq:spin_bath})), the expression for the temperature of the qubit at time $t$ is given by
 \begin{equation}
T_1(t)=\frac{E_1}{\log \left(\frac{(E_1^2+4) \left(e^{\frac{1}{{\tau_1}}}+1\right)\left(e^{\frac{E_1}{{\tau_1}}}+1\right)}{D}-1\right)},
 \end{equation}
 where $D=E_1^2\big(e^{\frac{1}{\tau_1}}+1\big)+2\cos\big(\sqrt{E_1^2+4t}\big)\big(e^{\frac{1}{\tau_1}}-e^{\frac{E_1}{\tau_1}}\big)+2\big(e^{\frac{1}{\tau_1}}+e^{\frac{E_1}{\tau_1}}\big)+4$. This temperature $T_1(t)$ exhibits an oscillatory behavior over time, ultimately reaching a temperature significantly lower than the initial temperature of the qubit (not shown in the paper). This behavior signifies a transient cooling of the qubit. If we now increase the number of spins of the non-Markovian environments, for instance, by setting $N=2$, the temperature exhibits a qualitatively similar behavior with time to that observed in the case with $N=1$. The behavior of temperature of the system qubit in the \(N=2\) case is as shown in Fig.~\ref{1q_2q}-(a). 
In this scenario, $T_1(t)$ oscillates uniformly between $1.0$ and $T_{1_{min}} \approx 0.5$, but the envelope of these oscillations is periodic and never reaches to equilibrium. So, there are instances of transient refrigeration at certain times. 
In contrast, when dealing with a Markovian environment, a single-qubit system cannot be cooled to a temperature lower than its initial value (which is assumed to be same as that of the environment). 
This provides us with the intuition that 
finite-size spin environments may play a crucial role in cooling, even in the three-qubit refrigerator scenario.

\subsection{Refrigeration with two-qubit 
two-environ setup}
\label{Aparajita}
Next we consider two two-level systems, each coupled to its respective 
environment locally. As in the case of three-qubit 
three-environment cooling process,  
 our objective here is to cool the first qubit and model a self-contained cooling process involving two qubits.  
 Initially, we assume that the systems are in equilibrium with their respective local environments, both maintained at the same temperatures, i.e., 
$\tau_1=\tau_2$. The Hamiltonian of the composite two-qubit system is $H_{S_1}+H_{S_2}$.  
The eigenvectors associated with this system Hamiltonian are $\ket{00}$, $\ket{01}$, $\ket{10}$, and $\ket{11}$ corresponding to the energy eigenvalues $\frac{K}{2}(E_1+E_2)$, $-\frac{K}{2}(E_2-E_1)$, $\frac{K}{2}(E_2-E_1)$, and $-\frac{K}{2}(E_1+E_2)$, respectively. In order to achieve cooling of the first qubit, our objective is to increase the population of the ground state of the first qubit. This can be accomplished by promoting transitions such as  $\ket{01} \leftrightarrow \ket{10}$ or $\ket{00} \leftrightarrow \ket{11}$, 
in which the probability of the forward flip is higher. 
For $E_2>E_1$, the probability of obtaining $\ket{01}$ is greater than that of $\ket{10}$, and 
the state $\ket{11}$ is 
more probable than $\ket{00}$. 
Hence, the transition from $\ket{00}$ to $\ket{11}$ is less probable, whereas by applying a suitable unitary transformation, we can increase the transition probability from $\ket{01}$ to $\ket{10}$.
This transition requires external energy because the energies of the states 
$\ket{01}$ and $\ket{10}$ 
are unequal
~\cite{Popescu}. Therefore, when $E_2>E_1$, 
the two-qubit 
cooling process is not self-contained. 
Now, to make the two-qubit 
cooling process as a self-contained one, we put the \textbf{condition (i) $E_1=E_2$} and turn on the interaction between the systems as 
$H_{\text{int}}^{1}=K g (\ketbra{01}{10}+\ketbra{10}{01})$, which reduce the energy cost to zero for the transition $\ket{01} \leftrightarrow \ket{10}$. The same objective can be attained with the \textbf{condition (ii) $E_1=-E_2$}, with the interaction $H_{\text{int}}^{2}=K g(\ketbra{00}{11}+\ketbra{11}{00})$ for the transition $\ket{00} \leftrightarrow \ket{11}$. Under Markovian environments,  neither of these two conditions can lead us to achieve transient or steady-state cooling in a refrigerator. Instead, these conditions lead to transient and steady-state heating of the cold qubit. We now take a two-qubit 
two-environment setup, where each of the qubits is connected to non-Markovian finite-size spin environments. 
In other words, the environment configuration is  $\{B^{NM}_1,B^{NM}_2\}$. We numerically obtained that under the self-contained condition (i), cooling of the first qubit is achievable when $E_1=E_2 <1.0$. Similar to the single qubit scenario, the temperature of the first qubit undergoes periodic oscillations, ultimately reaching a temperature significantly lower than the qubit's initial temperature at several instances of time. However, in contrast to the single qubit case, the cold qubit's temperature oscillates with a reduced amplitude. See Fig.~\ref{1q_2q}-(b). 
For condition (ii), there again exist instances of cooling 
of the first qubit.
Therefore, two-qubit self-contained cooling process 
can exist in presence of non-Markovian spin environments.
At certain times, however, the excited state population of the first qubit  becomes higher than that of its ground state, and therefore the definition of temperature breaks down.

\subsection{Refrigeration with  three-qubit 
three-environ setup}
\label{ekhane}

We now move on to the three-qubit 
three-environment 
cooling process discussed previously. We have already examined the impact of Markovian environments on the refrigeration process in the previous section. In this section, we consider a scenario where all three 
environments are 
finite-size spin environments, i.e., the 
environ configuration is $\big\{B_1^{NM},B_2^{NM},B_3^{NM}\big\}$. The time dynamics of the cold qubit temperature for this case
is depicted in Fig.~\ref{1q_2q}-(c). Like the single and the two-qubit cases with non-Markovian finite-size spin environments, here also the cold qubit temperature, denoted as $T_1(t)$, exhibits an oscillatory nature. The minimum transient temperature attained by qubit-$1$ is $< 0.82$. Although this transient cooling is not as effective as the single and two-qubit cases shown in Figs.~\ref{1q_2q}-(a) and (b), it still offers advantages over certain Markovian setups. For instance, this transient temperature is lower than the transient temperatures $\overline{T}^C$ and $T^{\prime C}$ in the ideal Markovian setups $\mathbf{S_1}$ and $\mathbf{S_2}$, respectively, as well as lower than the steady-state temperature  $T^{\prime S}$ in $\mathbf{S_2}$. However, it does not fall below the steady-state temperature $T^{\prime\prime S}$ in $\mathbf{S_3}$.
Therefore, we can infer that a significant degree of transient cooling of the first qubit can be attained with three non-Markovian environments.
It is important to note that while the oscillation rate of $T_1(t)$ decreases over time, a steady state is not reached due to the finite-size effect of the spin environments. 
In all the demonstrations of Fig.~\ref{1q_2q} we have set $N=2$.

\section{When only a subset of the 
environments are 
finite-size spin environments}
\label{m_nm}

Given that natural environments are often likely to exhibit non-Markovian behavior, the existence of self-contained cooling processes 
for single and two qubits, and the ability to retain the 
cooling capacity of the 
three-qubit setup 
in the presence of non-Markovian finite-size spin environments, 
 as seen in the preceding section, are interesting.
However, one may require a 
cooling process involving hybrid systems (like atom-photon ones) which may be prone to having some Markovian and other non-Markovian 
environments. Also,
certain non-Markovian 
environments can have Markovian limits. Hence, it is justifiable to consider a situation where among the two or the three 
environments of a two- or three-qubit quantum 
cooling device, respectively, 
there exists a combination of Markovian and non-Markovian 
environments.
The time evolution of an $(m+n)$-qubit system as a whole in presence of a combination of local Markovian and non-Markovian environments is described by the  equation~\cite{Ghoshal},
\begin{eqnarray}
\label{QME_mixed}
    &&\frac{\partial \tilde{\rho}_s(t)}{\partial t} =\mathcal{L}^{\prime}(\tilde{\rho}_s(t))= -\frac{i}{K}[H_S,\tilde{\rho}_s(t)]\nonumber\\
    &&+ \frac{\hbar}{K}\sum_{i_m}\mathcal{D}_{M_{i_m}}(\tilde{\rho}_s(t))+ \frac{\hbar}{K}\sum_{i_n}\mathcal{D}_{NM_{i_n}}(\rho_{sB_{\{i_n\}}^{NM}}(t)),\;\;\;\;
\end{eqnarray}
with $m$ and $n$ being the total number of Markovian and non-Markovian 
environments, respectively. Here, $i_m$ runs over all the qubits attached to Markovian environments and 
$i_n$ runs over the same coupled to 
non-Markovian ones. For example, if $m=1$ and $n=2$, and the first and third qubits are connected to non-Markovian 
environments while the second qubit is attached to a Markovian environment, then $i_m$ would be $2$ and the set $\{i_n\}$ would be given by $\{i_n\}=\{1,3\}$. 
The reduced state of the system $\tilde{\rho}_s(t)$ at time $t$ is given by $\tilde{\rho}_s(t)=\text{Tr}_{B_{\{i_n\}}}\big(\rho_{sB_{\{i_n\}}^{NM}}(t)\big)$, where $\rho_{sB_{\{i_n\}}^{NM}}(t)$ is the system's state correlated with all the non-Markovian 
environments. $\mathcal{D}_{M_{i_m}}(\tilde{\rho}_s(t))$ is the dissipative term, and appears for the presence of the $m^{\text{th}}$ Markovian environment, 
having the same form as in Eq.~(\ref{Lindblad}). The decay constants $\gamma_{i_m}(\omega^{\prime})$ and the Lindblad operators $L_{i_m}^{\omega^{\prime}}$ of this dissipative term for the corresponding $i_m$ is given by Eq.~(\ref{eq:A1}) in Appendix~\ref{appen:1}. $\mathcal{D}_{NM_{i_n}}(\rho_{sB_{\{i_n\}}^{NM}}(t))$ is the same coming from the 
qubit-environment interaction, for which the 
environ is a non-Markovian one. This term can be written as
\begin{equation}
    \mathcal{D}_{NM_{i_n}}(\rho_{sB_{\{i_n\}}^{NM}}(t)) = -\frac{i}{\hbar} \text{Tr}_{B_{\{i_n\}}^{NM}}\Big[ H_{SB_{i_n}^{NM}},\rho_{sB_{\{i_n\}}^{NM}}(t) \Big].
\end{equation}
In the case of this combined evolution involving both local Markovian and non-Markovian dynamics, the reduced subsystem of the first qubit again remains diagonal in the eigenbasis of the system Hamiltonian  $H_{S_1}$. Hence, the definition of local temperature, as provided in Eq.~(\ref{temp}), remains valid in these cases as well. 
We will now investigate the impacts of this combined Markovian and non-Markovian evolution on the 
cooling capacity of two- and three-qubit quantum absorption refrigerators. In the subsequent discussions, we will refer to the cooling process setups involving a combination of Markovian and non-Markovian 
environment as ``altered" setups. 
In the case of a two-qubit 
two-environ cooling process, there are two altered situations. These altered setups are characterized by the 
environment configurations $\mathcal{Q}_1=\big\{B_1^{M},B_2^{NM}\big\}$ and $\mathcal{Q}_2=\big\{B_1^{NM},B_2^{M}\big\}$. For three-qubit 
three-environment 
cooling devices, we have the three ideal Markovian setups $\mathbf{S_1}$, $\mathbf{S_2}$, and $\mathbf{S_3}$. See end of Sec.~\ref{Sec:2}. We now 
replace one or two of the Markovian environments of the ideal setups 
with non-Markovian ones. This approach yields four modified scenarios for each of $\mathbf{S_1}$, $\mathbf{S_2}$, and $\mathbf{S_3}$. For $m=2$ and $n=1$, i.e., two Markovian and one non-Markovian environments, we take two 
environment configurations: $\mathcal{A}_1^{S_i}\equiv \big\{B_1^{NM},B_2^M,B_3^M\big\}$ and $\mathcal{A}_2^{S_i} \equiv \big\{B_1^{M},B_2^M,B_3^{NM}\big\}$. For $m=1$ and $n=2$, we consider two scenarios with the environment configurations: $\mathcal{A}_3^{S_i} \equiv \big\{B_1^{NM},B_2^M,B_3^{NM}\big\}$ and $\mathcal{A}_4^{S_i} \equiv \big\{B_1^{M},B_2^{NM},B_3^{NM}\big\}$. We remember that the qubit 1 is the ``cold'' qubit. The superscripts $S_i$ denote the respective ideal situations $\mathbf{S_1}$, $\mathbf{S_2}$ and $\mathbf{S_3}$ for $i=1,2$ and $3$, respectively. In all four altered situations of $\mathbf{S_1}$, $\mathbf{S_2}$, and $\mathbf{S_3}$, the parameters of the Markovian environments, denoted as $\alpha_i$ for $i=1, 2$, and $3$, are the same as in the ideal $\mathbf{S_1}$, $\mathbf{S_2}$, and $\mathbf{S_3}$ scenarios, respectively. Additionally, all the parameters of the non-Markovian environments remain consistent with those in Fig.~\ref{1q_2q}-(c). Therefore, the parameter spaces corresponding to these four scenarios $\mathcal{A}_1^{S_i}$, $\mathcal{A}_2^{S_i}$, $\mathcal{A}_3^{S_i}$ and $\mathcal{A}_4^{S_i}$ are respectively $\big\{\alpha_0,\alpha_2^{S_i},\alpha_3^{S_i}\big\}$, $\big\{\alpha_1^{S_i},\alpha_2^{S_i},\alpha_0\big\}$, $\big\{\alpha_0,\alpha_2^{S_i},\alpha_0\big\}$ and $\big\{\alpha_1^{S_i},\alpha_0,\alpha_0\big\}$. It is important to note that the scenarios $\mathcal{A}_1^{S_1}$, $\mathcal{A}_3^{S_1}$ and $\mathcal{A}_4^{S_2}$ are same as $\mathcal{A}_1^{S_2}$, $\mathcal{A}_3^{S_2}$ and $\mathcal{A}_4^{S_3}$, respectively, since $\alpha_1^{S_2}=\alpha_1^{S_3}$, $\alpha_2^{S_3}=\alpha_2^{S_2}$, and $\alpha_3^{S_1}=\alpha_3^{S_2}$. We remember that the \(\alpha_i\) are involved in the Ohmic spectral densities corresponding to interaction strength. See Eq.~(\ref{mark_int}). The different cases are tabulated in Fig.~\ref{tabular}. 
\subsection{Cooling 
in  altered situations $\mathcal{Q}_1$ and $\mathcal{Q}_2$}

\begin{figure}
\includegraphics[width=8cm,height=5.5cm]{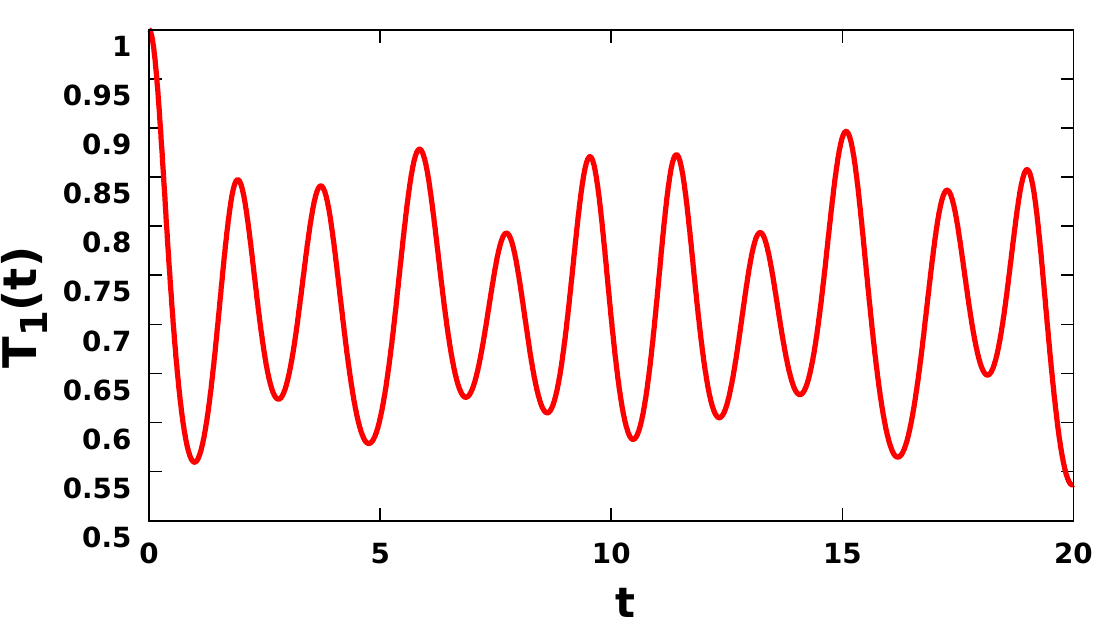}%

    \caption{Effects of combination of local environments on a two-qubit quantum refrigerator, which is self-contained under the condition $E_1=E_2$. The environment configuration is $\mathcal{Q}_2=\big\{B_1^{NM},B_2^{M}\big\}$, $N=2$, and $\alpha_2=10^{-4}$. Here we have taken $E_1=E_2=1/2$, and the temperatures of the environments are 
 $\tau_1=\tau_2=1.0$. All quantities plotted here are dimensionless.}
    \label{two_1m_1nm}
\end{figure}

For the configuration $\mathcal{Q}_1$ with the self-contained conditions (i) and (ii) for the two-qubit quantum absorption refrigerator (see Sec.~\ref{Aparajita}),
it is observed that when $\tau_1 \le \tau_2$, the cold qubit temperature $T_1(t)$ always remains higher than $\tau_1$ at every instant of time, resulting in no cooling. 
On the other hand, for the configuration $\mathcal{Q}_2$, with the self-contained conditions (i) and (ii), we find 
instances of cooling. 
See Fig.~\ref{two_1m_1nm}. The temperature of the cold qubit oscillates rapidly in the transient regime and reaches the minimum temperature $\approx 0.55$ at certain instances of time.  

\begin{figure}
\includegraphics[width=8.6cm,height=6.5cm]{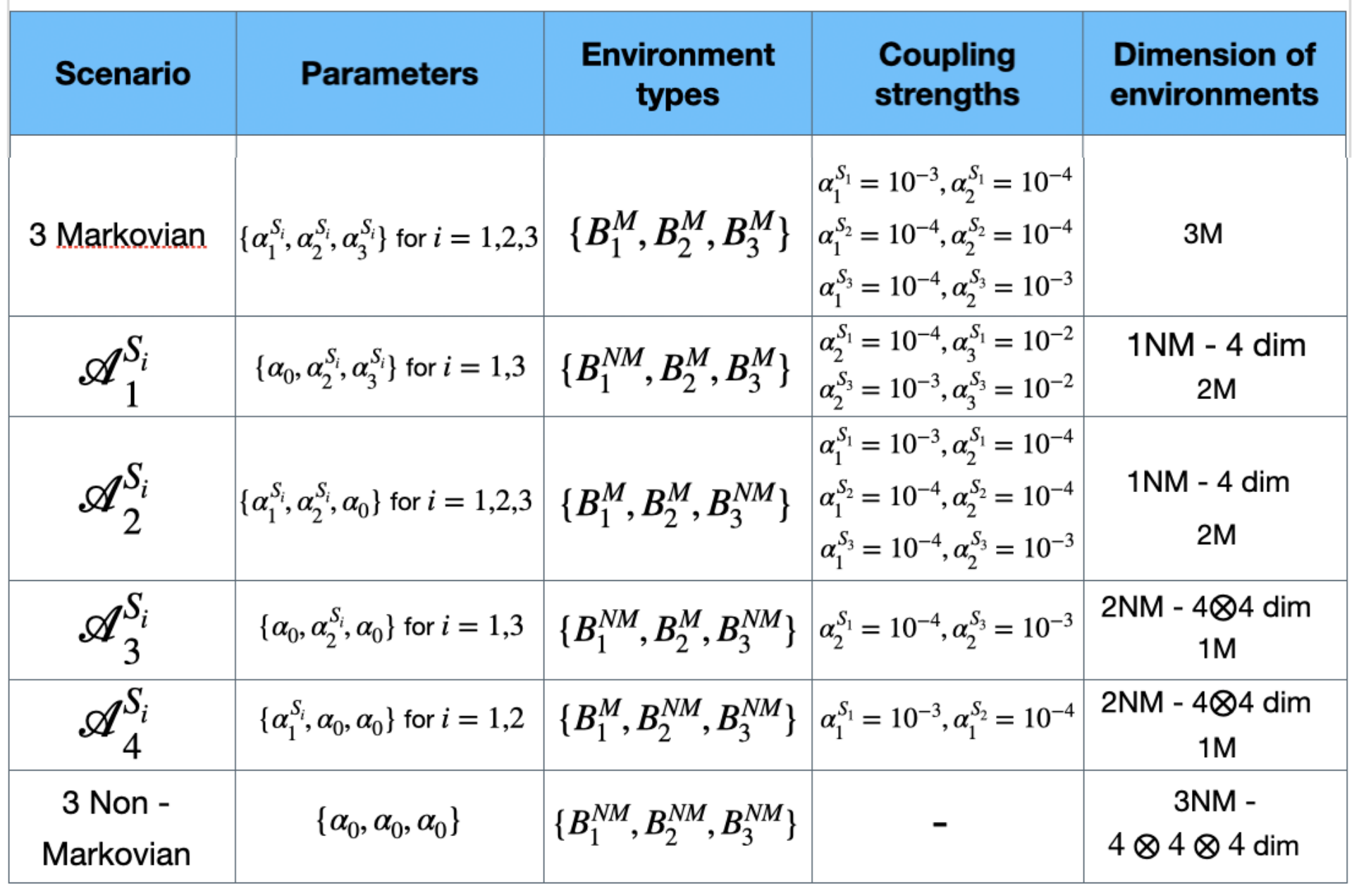}%

    \caption{We present here a summary of the different scenarios considered in the paper, and point to the relevant parameters, environment types, coupling strengths and dimensions of the environments. The initial states chosen are the thermal states for each of the qubits and the spin-environments. The value of the coupling strength between the spin and environment is chosen to be $\alpha=0.5$ throughout the paper. The Markovian environments comprise infinite number of harmonic oscillators. The abbreviations, $M$ and $NM$ denote the Markovian and non-Markovian (finite-size environment) respectively.}
    \label{tabular}
\end{figure}

\subsection{Cooling in altered situation $\mathcal{A}_1^{S_i}$}
\label{TC}
\begin{figure*}
\centering
\includegraphics[width=8cm]{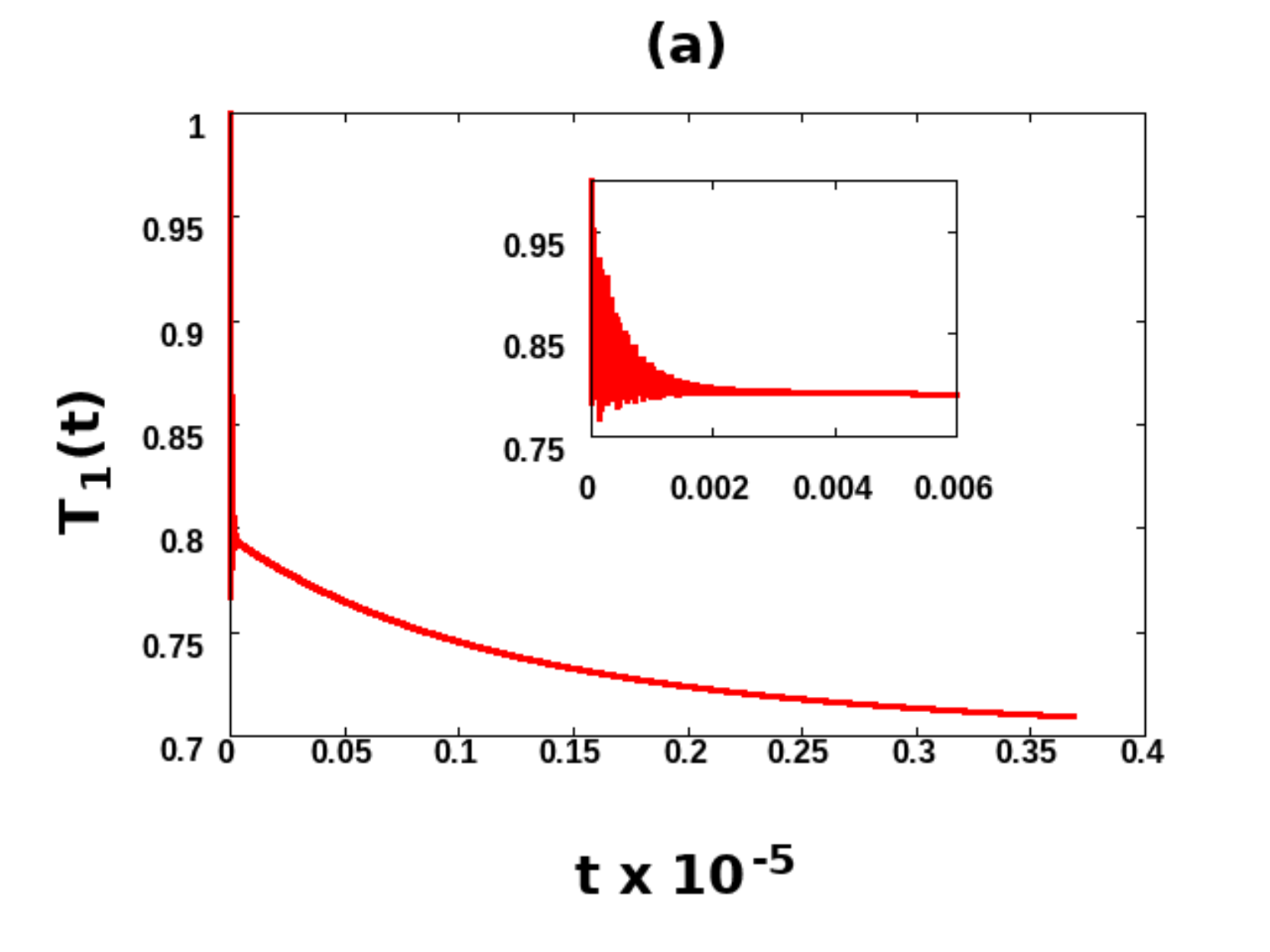}%
\includegraphics[width=8cm]{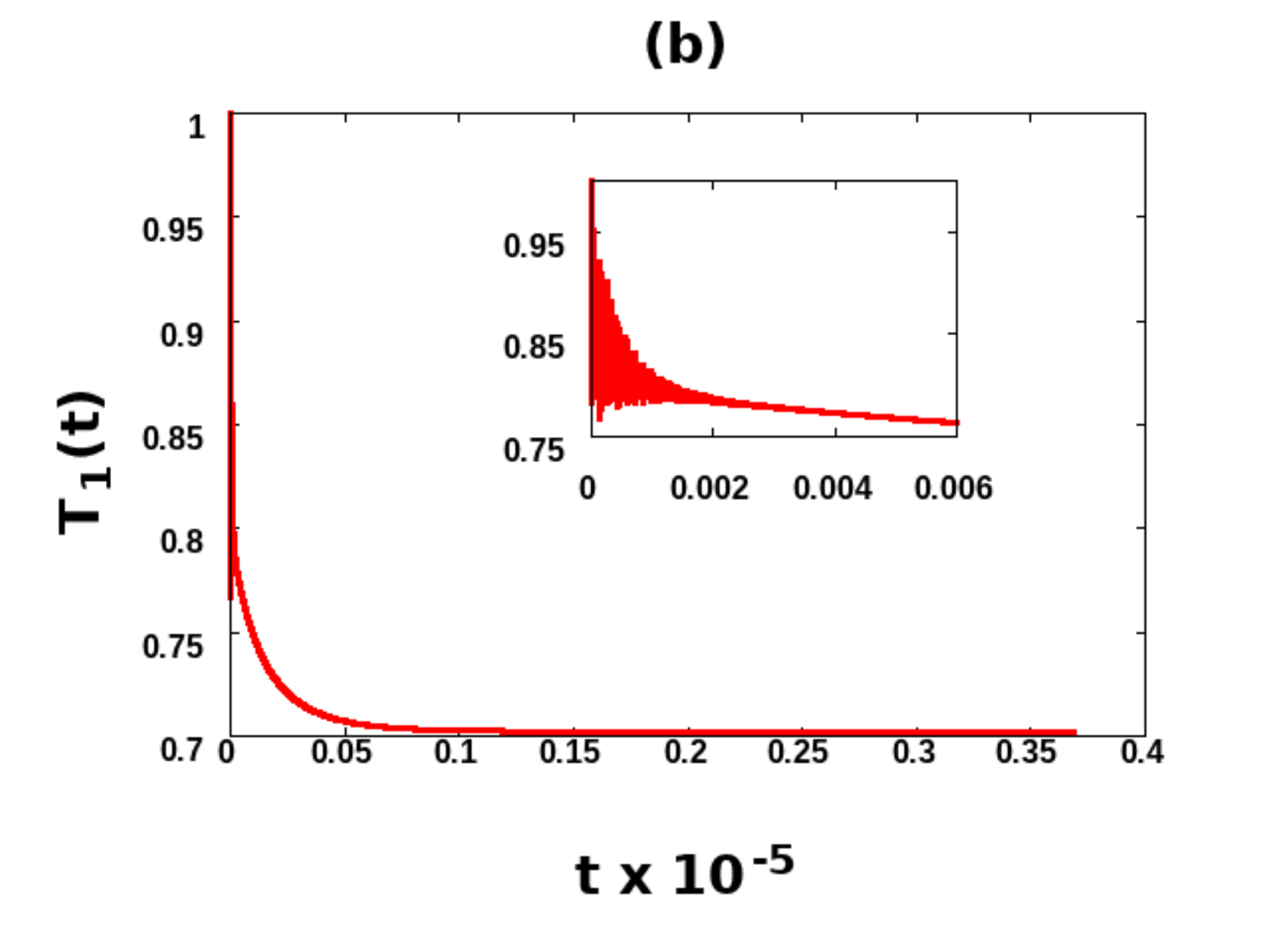}%
\caption{Effects of combination of local environments on cooling of the three-qubit refrigerator setup for the altered situations $\mathcal{A}_1^{S_i}$ with the parameter regime $\big\{\alpha_0,\alpha_2^{S_i},\alpha_3^{S_i}\big\}$. Here we depict
the dependence of $T_1(t)$ with $t$ 
for (a) $\mathcal{A}_1^{S_1}$ with $\big\{\alpha_0,\alpha_2^{S_1},\alpha_3^{S_1}\big\}$, (b) $\mathcal{A}_1^{S_3}$ with $\big\{\alpha_0,\alpha_2^{S_3},\alpha_3^{S_3}\big\}$. The configuration $\mathcal{A}_1^{S_2}$ is the same as the situation $\mathcal{A}_1^{S_1}$, since $\alpha_2^{S_2}=\alpha_2^{S_1}$ and $\alpha_3^{S_2}=\alpha_3^{S_1}$. 
The insets in panels (a) and (b) present  magnified versions of the transient behavior near the initial time. The system parameters and the specifications for each of the Markovian and non-Markovian environments are taken to be the same as in Figs.~\ref{3mark} and~\ref{1q_2q}-(c), respectively. All  quantities plotted along both the axes are dimensionless.}
\label{1st_nm}
\end{figure*}

\begin{figure}
\centering
\includegraphics[width=8cm]{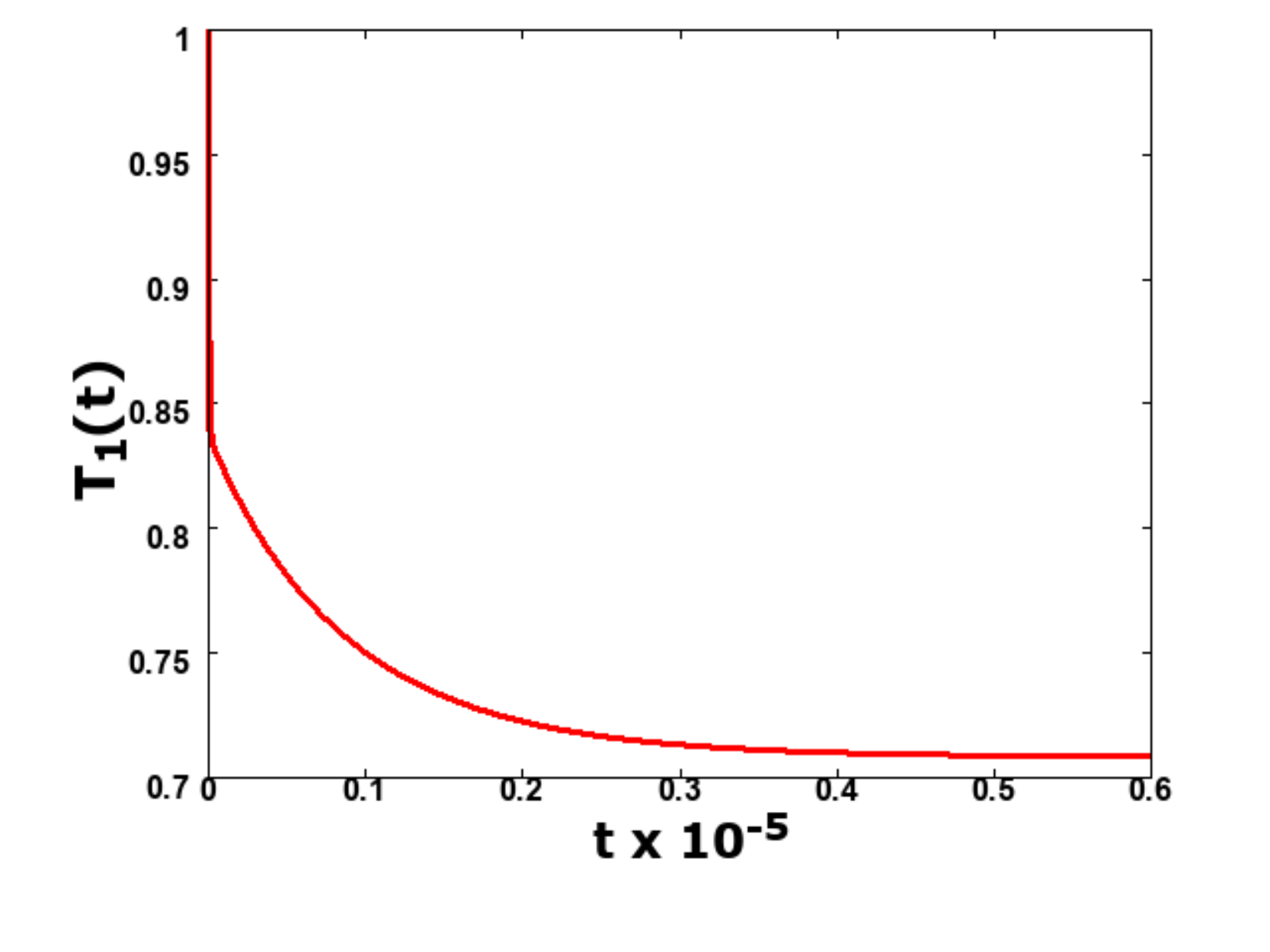}%
\caption{Time dynamics of the cold qubit temperature of a three-qubit three-environment model of a quantum absorption refrigerator when the second and third qubits are attached separately to Markovian environments. Here we depict the dependence of $T_1(t)$ on $t$ in the operating regime  $\mathbf{S_1}$, i.e.  $\alpha_1=10^{-3}$, $\alpha_2=10^{-4}$, $\alpha_3=10^{-2}$. All  quantities plotted along both the axes are dimensionless.}
\label{1nospin}
\end{figure}

Let us now move over to the three-qubit cases. Consider the scenario where a three-qubit ideal quantum absorption refrigerator operating in the parameter regime $\mathbf{S_1}$ 
is altered by replacing the Markovian environment attached to the cold qubit by a non-Markovian 
spin-environment, giving us the configuration $\mathcal{A}_1^{S_1}$, i.e., $\big\{B_1^{NM},B_2^M,B_3^M\big\}$ with the parameters $\{\alpha_0,\alpha_2^{S_1},\alpha_3^{S_1}\}$. The system-refrigerator interaction is now governed by $H_{SB}=H_{SB_1^{NM}}+\sum_{i=2}^3 H_{SB_i^M}$. 
In Fig.~\ref{1st_nm}-(a) the dynamics of  temperature of the first qubit for this scenario is depicted. We observe that
the refrigeration capacity of the device is sustained. The transient temperature $T_1(t)$ approaches a minimum $\approx 0.77$, which is less than the same of the corresponding ideal Markovian case $\overline{T}_1^C$. In course of time, the cold qubit temperature slowly attains a steady state temperature $\approx 0.7$. 
Hence, it is clear that a steady state is reached when only one Markovian bath is replaced by a spin environment. The reachability of steady states when more than one Markovian (infinite) bath is present occurs because the influence of the infinite Markovian baths dominates the dynamics of the qubits, leading to steady-state behavior. Steady-state would still be achieved even if the cold qubit were not connected to any bath, as long as the other two qubits are coupled to Markovian baths. See Fig.~\ref{1nospin}. 

If we now consider the parameter regime, $\mathbf{S_2}$, and replace the first Markovian bath by a non-Markovian one, we get the configuration $\mathcal{A}_1^{S_2}$. This configuration is the same as $\mathcal{A}_1^{S_1}$, as $\alpha_2$ and $\alpha_3$ are equal in these cases.

We now look at the effect of combined local Markovian and non-Markovian evolution on the 
cooling process with the combination of the environments being $\big\{B_1^{NM},B_2^M,B_3^M\big\}$, but in the parameter space $\mathbf{S_3}$, i.e., with the parameters  $\{\alpha_0,\alpha_2^{S_3},\alpha_3^{S_3}\}$.
As in the ideal setup (shown in Fig.~\ref{3mark}-(c)), in this case also,  SSC is better than  TC 
with the steady-state temperature $<T_1^{{\prime\prime}^S}$, where $T_1^{{\prime\prime}^S}$ is the steady-state temperature for the ideal setup $\mathbf{S_3}$. Compare Figs.~\ref{1st_nm}-(b) and~\ref{3mark}-(c). 
Again, the attainment of steady state is because of the presence of the two Bosonic Markovian baths connected to the second and third qubits.
Hence, compared to the ideal setup $\mathbf{S_3}$, an enhancement of cooling occurs in this scenario 
in the transient regime.

Analysis of the above altered situations tell us  that replacement of the Markovian bath attached to the cold qubit of an ideal absorption refrigerator by a non-Markovian one in the operating regime $\mathbf{S_1}$ (or $\mathbf{S_2}$),  enhances transient cooling by reducing the temperature of the cold qubit below the cooling obtained in the ideal Markovian case. I.e., the non-Markovian cooling process leads us to attain temperatures $< \overline{T}_1^C$ and $<T^{\prime^C}_1$ in regimes \(S_1\) and \(S_2\). (See end of Sec.~\ref{Sec:2}.) 
Similarly, substituting the first Markovian reservoir by a non-Markovian one for the setting  $\mathbf{S_3}$, one attains an improved refrigeration both in the transient regime. 
Hence, the results obtained here reveal that if we substitute the Markovian bath attached to the cold qubit by a non-Markovian one (say a spin environment), both transient and steady state coolings can be achieved, which can often be better than certain ideal Markovian setups. 

\subsection{Cooling in  altered situation $\mathcal{A}_2^{S_i}$}
\label{TC}
\begin{figure*}
\centering
\includegraphics[width=5.8cm]{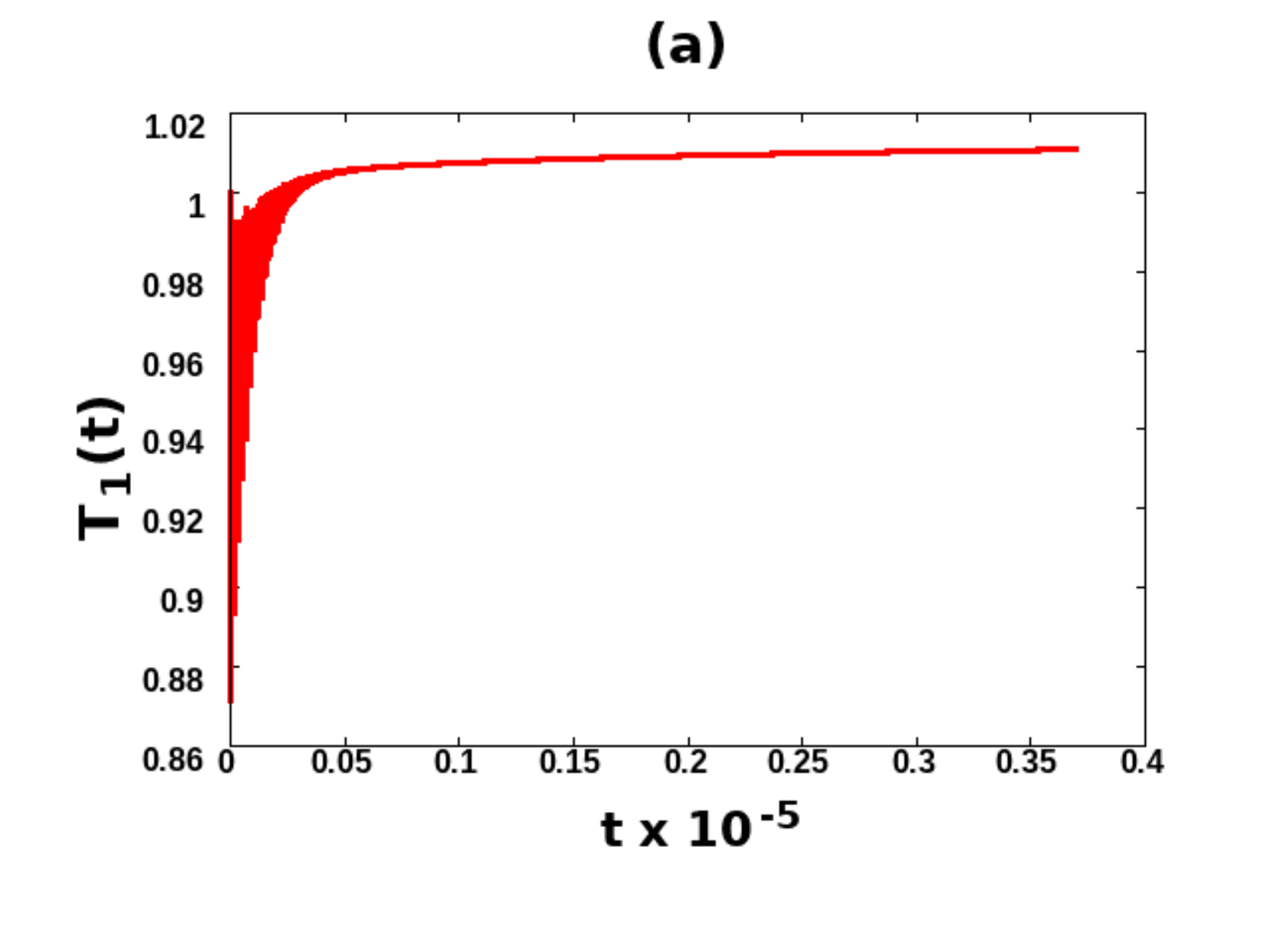}%
\hspace{.25cm}%
\includegraphics[width=5.8cm]{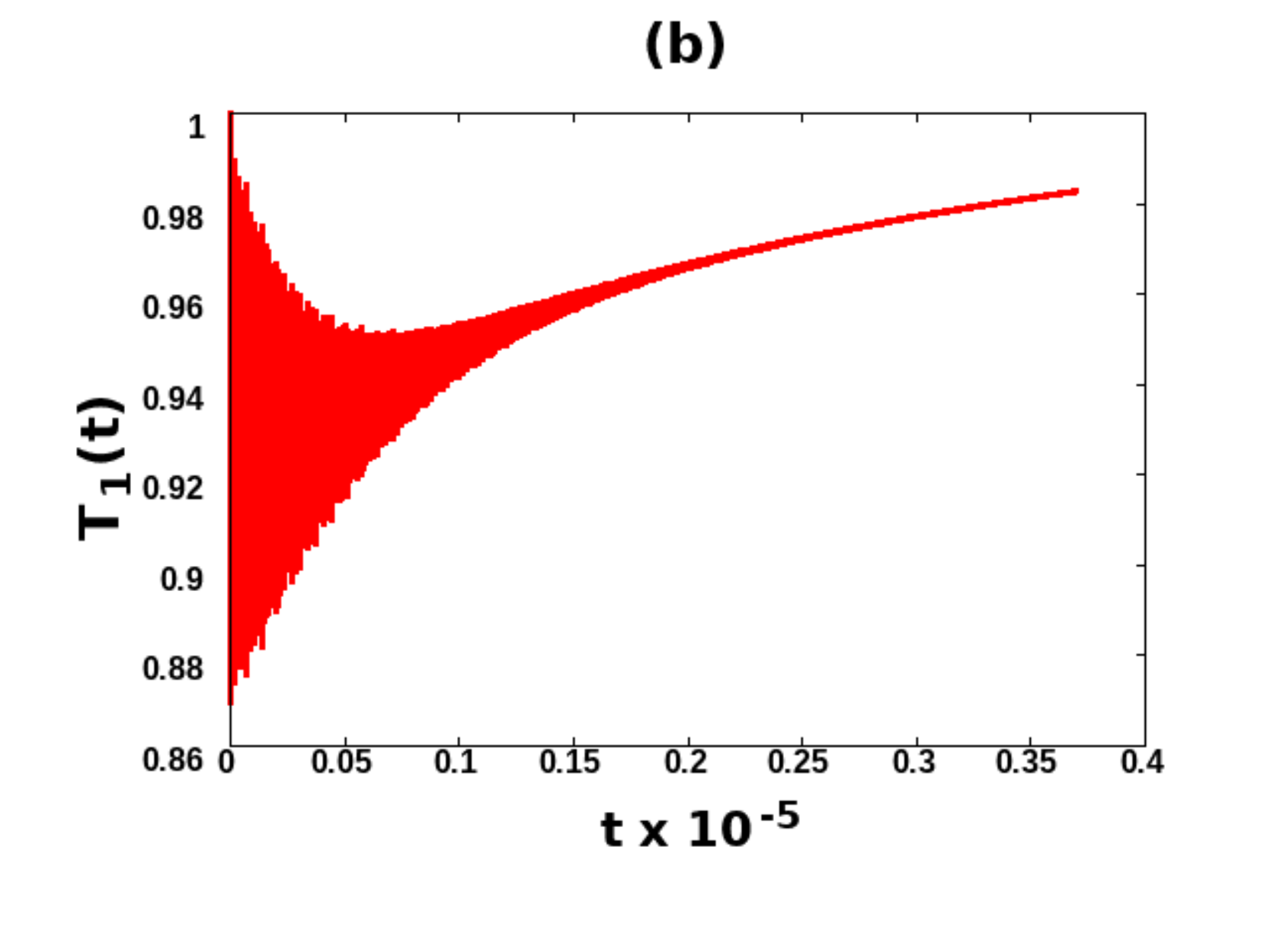}%
\hspace{.25cm}%
\includegraphics[width=5.8cm]{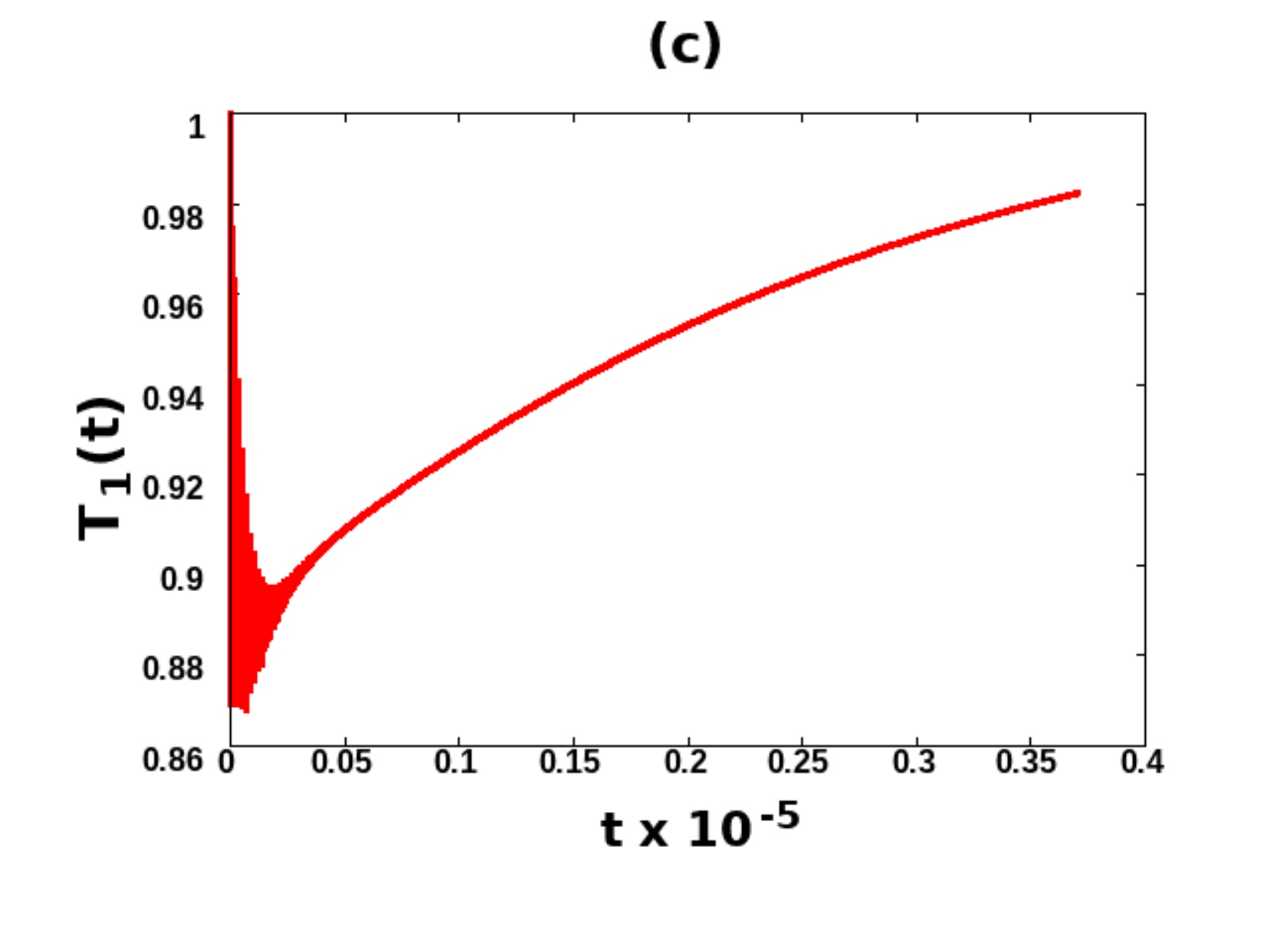}%
\caption{Effects of combination of local environments on cooling of the three-qubit refrigerator setup for the altered situations $\mathcal{A}_2^{S_i}$ with the parameter regime $\big\{\alpha_1^{S_i},\alpha_2^{S_i},\alpha_0\big\}$. Here we 
plot the dependence of $T_1(t)$ with $t$ by replacing the third Markovian bath with a non-Markovian one for 
(a) $\mathcal{A}_2^{S_1}$ with $\big\{\alpha_1^{S_1},\alpha_2^{S_1},\alpha_0\big\}$, (b) $\mathcal{A}_2^{S_2}$ with $\big\{\alpha_1^{S_2},\alpha_2^{S_2},\alpha_0\big\}$ and (c) $\mathcal{A}_2^{S_3}$ with $\big\{\alpha_1^{S_3},\alpha_2^{S_3},\alpha_0\big\}$. 
The system parameters and the specifications for each of the Markovian and non-Markovian environments are taken to be the same as in Figs.~\ref{3mark} and~\ref{1q_2q}-(c), respectively. All quantities demonstrated along both the axes are dimensionless.}
\label{3rd_nm}
\end{figure*}

Consider again the ideal three-qubit scenario where all the three baths are Markovian, and then we substitute any one of the two baths, which is not connected to the cold qubit, by a non-Markovian bath. We choose the third Markovian bath to be replaced by a non-Markovian one and the bath configuration becomes
$\big\{B_1^{M},B_2^M,B_3^{NM}\big\}$.  
For the altered setup $\mathcal{A}_2^{S_1}$, corresponding to the ideal $\mathbf{S_1}$ scenario, we show that one can obtain transient cooling,  
 as evident from Fig.~\ref{3rd_nm}-(a). In this situation, the minimum transient temperature is $\approx 0.88$. 
%
Similarly, for the altered configuration of $\mathbf{S_2}$, if the third qubit is connected to the non-Markovian spin environment with the others being Markovian, i.e., for $\mathcal{A}_2^{S_2}$,  transient 
cooling is obtained but 
is not advantageous with respect to the ideal case.
See Fig.~\ref{3rd_nm}-(b). Here the minimum transient temperature is $\approx 0.88$, 
 which is greater than the minimum transient 
temperature in the corresponding ideal Markovian case $\mathbf{S_2}$. This feature is different from in the preceding subsection (i.e.,  the $\mathcal{A}_1^{S_i}$ scenario), where there was an advantage over the corresponding ideal Markovian situations.
Similarly, the configuration $\mathcal{A}_2^{S_3}$ exhibits transient cooling of the first qubit as depicted in Fig.~\ref{3rd_nm}-(c). The minimum transient temperature lies beneath $\approx 0.88$ and equilibrates to  a value $\approx 0.98$.
So in the altered setups, $\mathcal{A}_2^{S_i}$, the action of the refrigerator in the transient regime still persists without any advantage over the ideal Markovian cases.
%
Also, though there exist bounded oscillations in $T_1(t)$ 
in the transient region, the temperatures for all these three cases attain  a steady state for long times. 

\subsection{Cooling in  altered situation $\mathcal{A}_3^{S_i}$}

\begin{figure*}
\centering
\includegraphics[width=8cm]{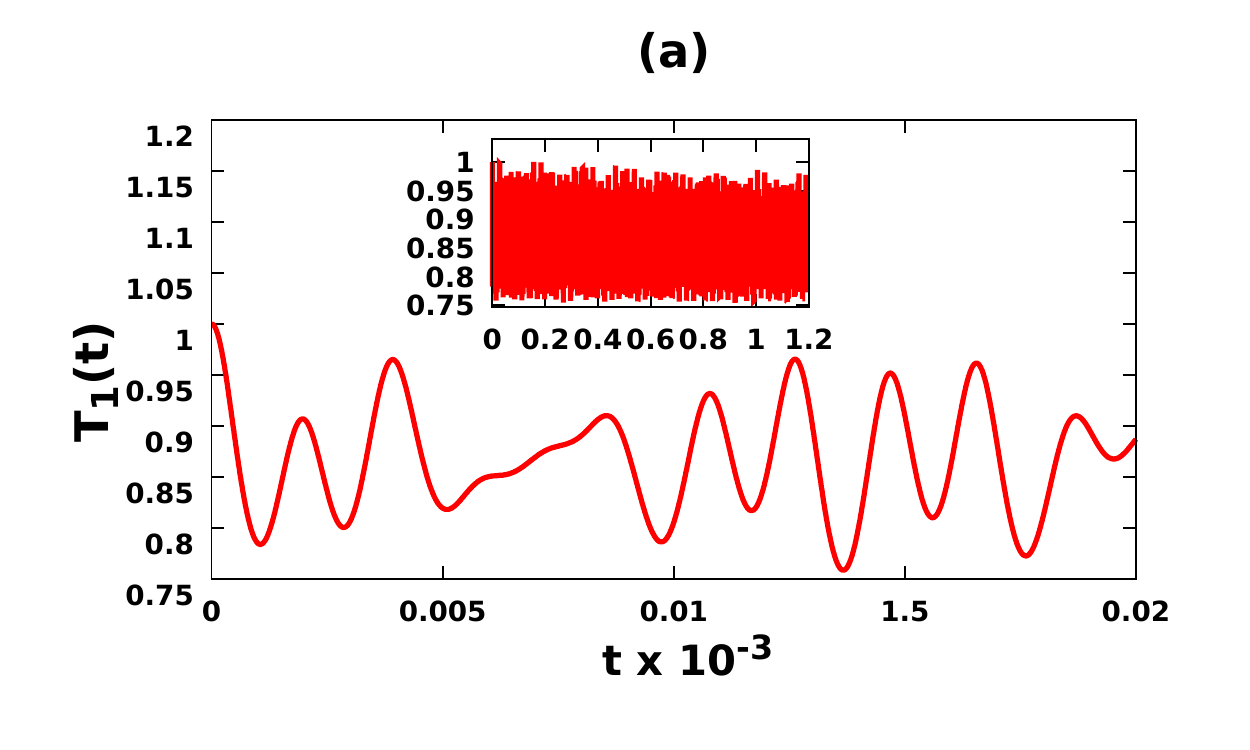}%
\includegraphics[width=8cm]{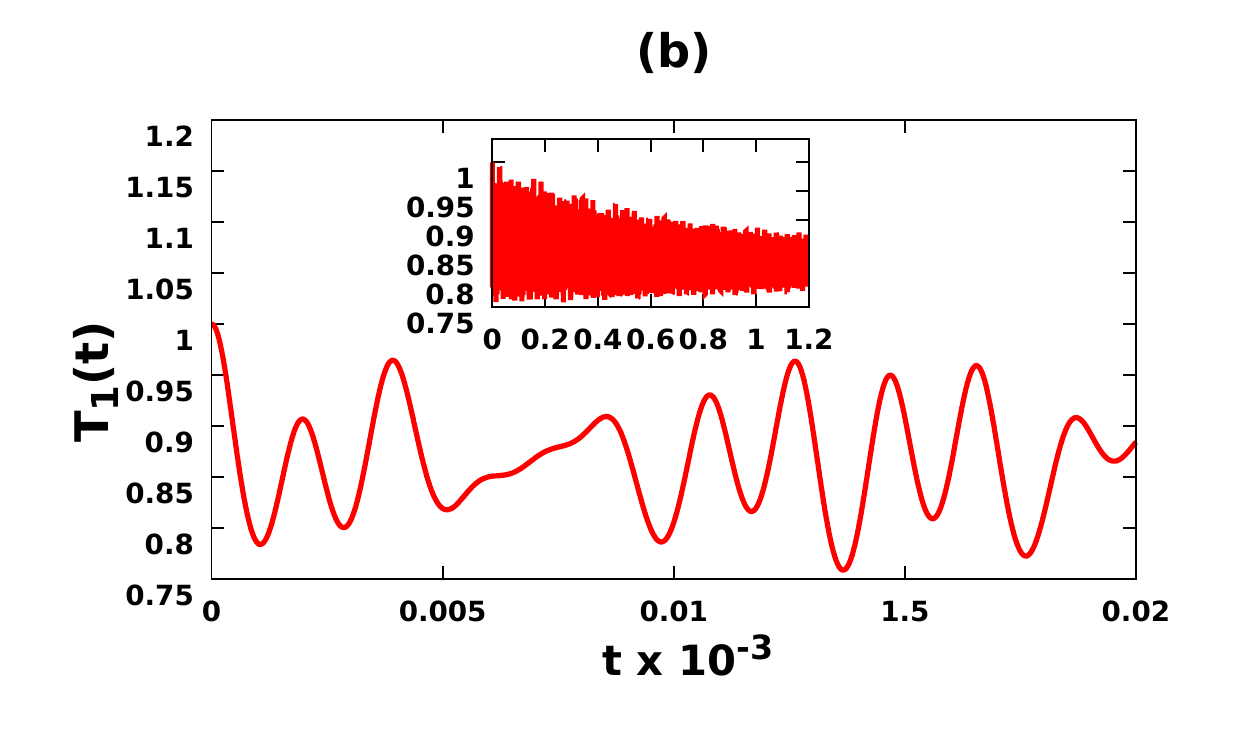}%
\caption{Effects of mixed set of local environments on cooling of the three-qubit refrigerator setup for the altered situations $\mathcal{A}_3^{S_i}$ with the parameter regime $\big\{\alpha_0,\alpha_2^{S_i},\alpha_0\big\}$. Here we depict the dependence of $T_1(t)$ with $t$ 
for (a) $\mathcal{A}_3^{S_1}$ with $\big\{\alpha_0,\alpha_2^{S_1},\alpha_0\big\}$ and (b) $\mathcal{A}_3^{S_3}$ with $\big\{\alpha_0,\alpha_2^{S_3},\alpha_0\big\}$.
The configuration $\mathcal{A}_3^{S_2}$ is the same as $\mathcal{A}_3^{S_1}$ since we take $\alpha_2^{S_2}=\alpha_2^{S_1}$.
The system parameters and the specifications for each of the Markovian and non-Markovian environments are taken to be the same as in Figs.~\ref{3mark} and~\ref{1q_2q}-(c) respectively. All quantities plotted here are dimensionless.}
\label{1st_3rd_nm}
\end{figure*}

Here we take an ideal three-qubit quantum refrigerator 
and replace two of the Markovian baths, say the first and the third one, by spin-environments.
So, the setup is now $\mathcal{A}_3^{S_i}$, i.e., the environment configuration is $\big\{B_1^{NM},B_2^M,B_3^{NM}\big\}$, where the system-environment interaction is governed by $H_{SB}=H_{SB_2^M}+\sum_{\{i\}} H_{SB_i^{NM}}$ with $\{i\}=\{1,3\}$.
Let us first look at the scenario $\mathcal{A}_3^{S_1}$ considered in Fig.~\ref{1st_3rd_nm}-(a). 
The transient temperature $T_1(t)$ oscillates rapidly between $0.75$ and $1.0$ with its magnitude never surpassing unity. The minima of $T_1(t)$ varies around $\approx 0.75$, which implies that along with obtaining  refrigeration, we also get  an enhancement in cooling of the first qubit as compared to the minimum transient temperature, $\overline{T}_1^C$, of the corresponding ideal Markovian case $\mathbf{S_1}$. Compare Figs.~\ref{3mark}-(a) and~\ref{1st_3rd_nm}-(a). 

Next we go over to the configuration $\mathcal{A}_3^{S_2}$. As we have previously mentioned, the situation $\mathcal{A}_3^{S_2}$ is the same as that of $\mathcal{A}_3^{S_1}$, and so this scenario mimics the one presented in Fig.~\ref{1st_3rd_nm}-(a). 
An important point to be noted here is that the attainment of equilibrium is non-existent in this scenario to within a significantly large time-scale. Here, the inability to reach steady states when fewer than two Markovian baths are present is a consequence of the finite size of the environments.

We now consider the parameters corresponding to the configuration, $\mathcal{A}_3^{S_3}$ (Fig.~\ref{1st_3rd_nm}-(b)), where we observe similar features as in $\mathcal{A}_3^{S_1}$ near the initial time. The transient temperature, $T_1(t)$, varies between $\approx 0.75$ and $1.0$, and the envelope of oscillations narrows down within the observation time.
There is significant  refrigeration obtained in this case, despite the oscillatory nature of the temperature profile, 
and the minimum transient temperature ($\approx 0.75$) is almost equal to the steady state temperature, $T_1^{\prime\prime S}$, of the corresponding ideal Markovian case $\mathbf{S_3}$. Compare with Fig.~\ref{3mark}-(c). Another configuration with $m=1$ and $n=2$ can be obtained by taking the altered situation $\big\{B_1^{NM},B_2^{NM},B_3^M\big\}$. We find that this configuration shows qualitatively the same results as $\big\{B_1^{NM},B_2^{M},B_3^{NM}\big\}$.
One distinct nature of these $m=1$, $n=2$ cases is that the oscillations of $T_1(t)$  persist for a significantly large time-scale. This behavior appears with the increase in the number of finite-size spin environments 
and qualitatively resembles the situation where all the three environments are non-Markovian. Compare with Fig.~\ref{1q_2q}-(c).
Therefore, substituting any two Markovian baths including the first one is beneficial as this enhances the transient cooling. 

\subsection{Cooling in  altered situation  $\mathcal{A}_4^{S_i}$}

\begin{figure*}
\centering
\includegraphics[width=8cm]{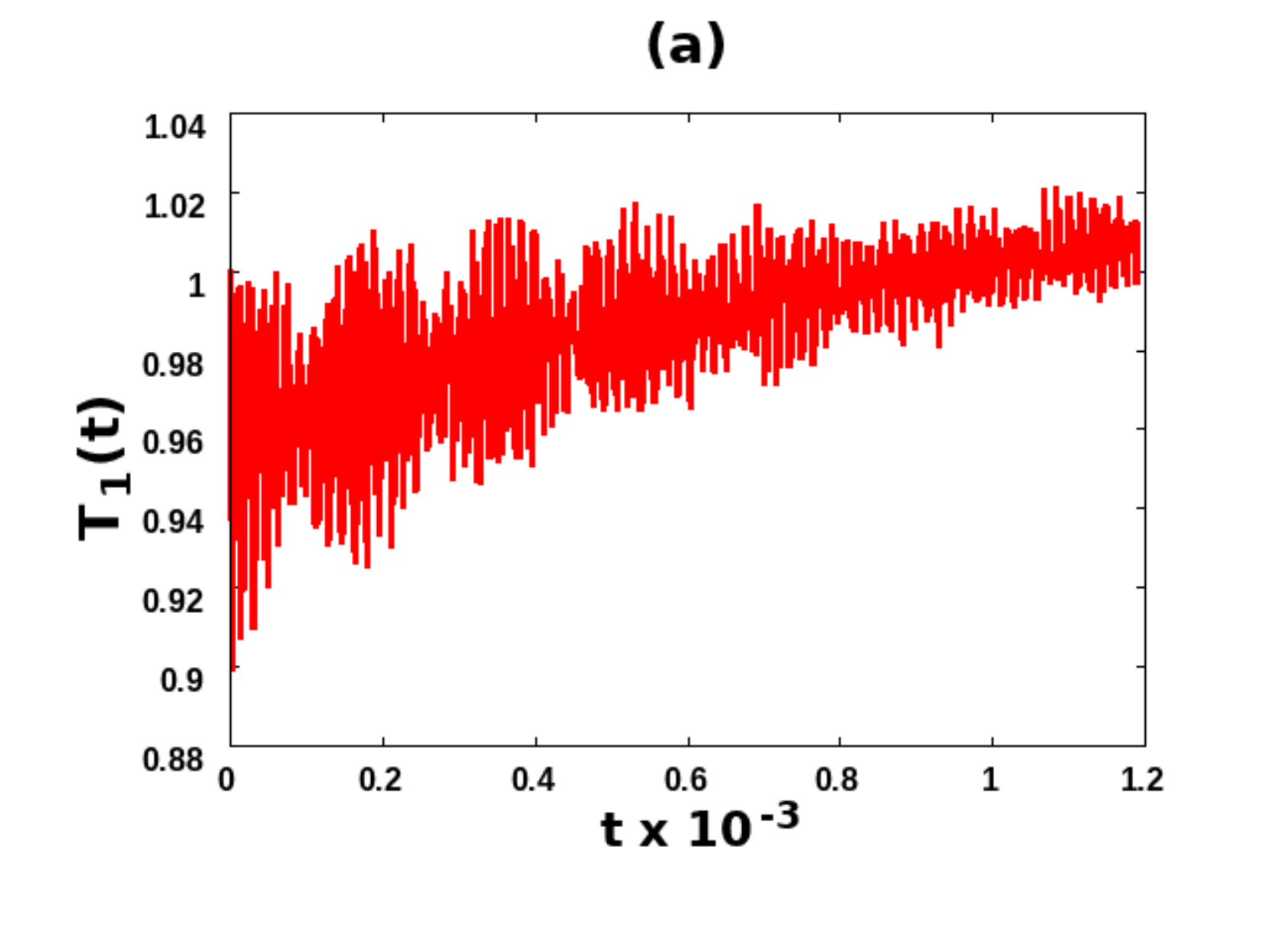}%
\includegraphics[width=8cm]{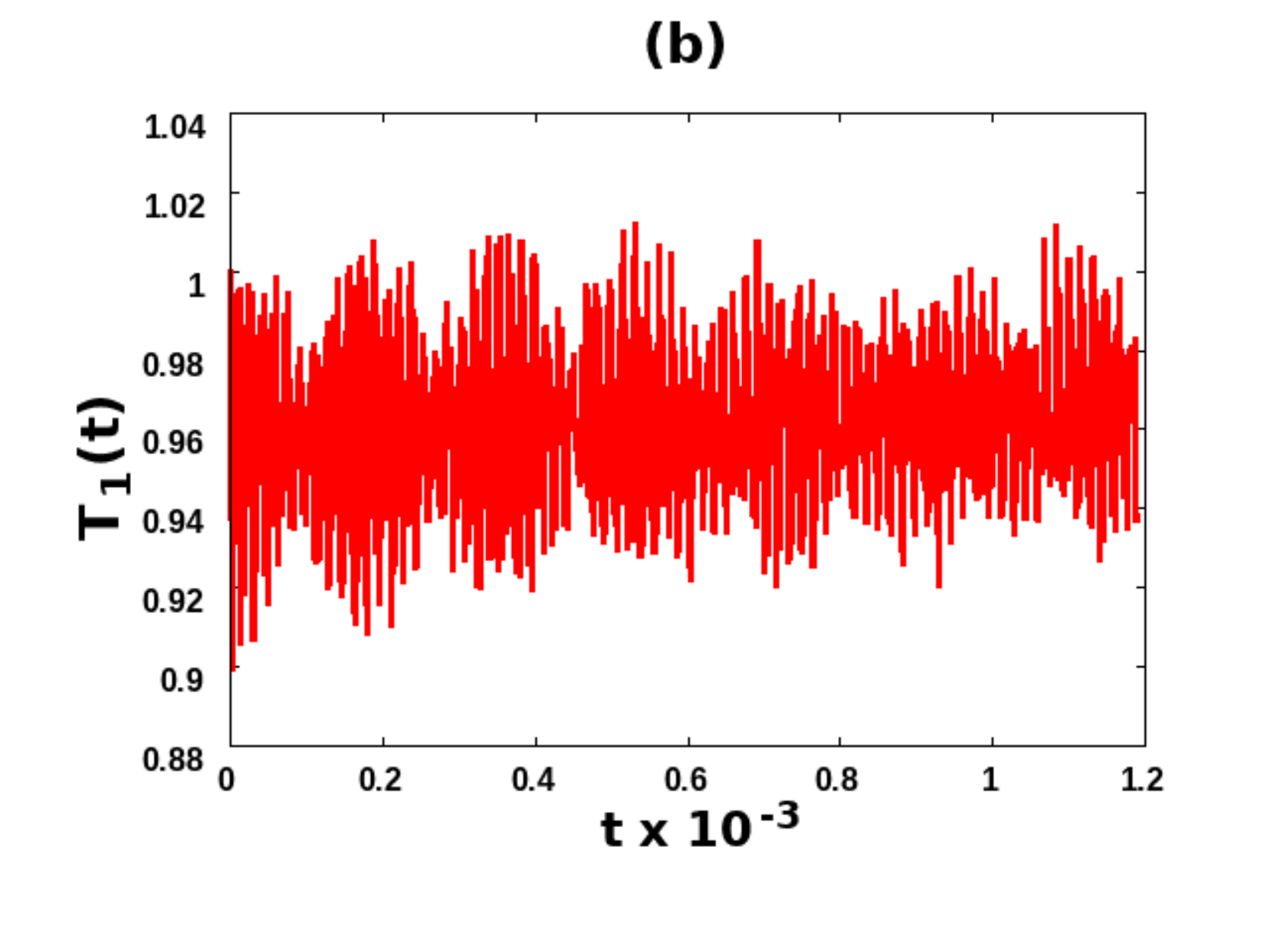}%
\caption{Effects of combination of local environments on cooling of the three-qubit refrigerator setup for the altered situations $\mathcal{A}_4^{S_i}$ with the parameter regime $\big\{\alpha_1^{S_i},\alpha_0,\alpha_0\big\}$. Here we demonstrate the dependence of $T_1(t)$ with $t$
for (a) $\mathcal{A}_4^{S_1}$ with $\big\{\alpha_1^{S_1},\alpha_0,\alpha_0\big\}$ and (b) $\mathcal{A}_4^{S_2}$ with $\big\{\alpha_1^{S_2},\alpha_0,\alpha_0\big\}$. 
The configuration of $\mathcal{A}_4^{S_2}$ matches with $\mathcal{A}_4^{S_3}$ for the chosen values of $\alpha_1$ in the ideal setups of $\mathbf{S_2}$ and $\mathbf{S_3}$.
The system parameters and the specifications for each of the Markovian and non-Markovian environments are the same as in Figs.~\ref{3mark} and~\ref{1q_2q}-(c) respectively. All quantities plotted here are dimensionless.}
\label{2nd_3rd_nm}
\end{figure*}

In this subsection, we again discuss a configuration of environments with $m=1$ and $n=2$, but different from the situations mentioned in the preceding subsections, we look here into the case where the second and third qubits are connected to  non-Markovian spin-environments while the first one is coupled to a Markovian one, i.e., the scenario is $\mathcal{A}_4^{S_i}$. We start with the altered $\mathbf{S_1}$ situation, $\mathcal{A}_4^{S_1}$. 
Unlike the previous instances of $\mathbf{S_1}$ where we had $m=1$ Markovian baths  and $n=2$ non-Markovian environments, we find 
that the device almost fails to refrigerate. See Fig.~\ref{2nd_3rd_nm}-(a). The temperature of the cold qubit $T_1(t)$ exhihits an oscillatory nature and the oscillations vary between the temperatures $0.9$ and $1.02$, with the minimum transient temperature being $\approx 0.9$. 


We now move over to the altered situation of $\mathbf{S_2}$, i.e. $\mathcal{A}_4^{S_2}$. For this setup (and the same occurs with $\mathcal{A}_4^{S_3}$), 
we observe that
the transient temperature, $T_1(t)$, of the cold qubit  begins to oscillate between $0.9$ and $< 1.02$. 
So here also, 
no enhancement of the transient state cooling is achieved over the ideal Markovian situation compared to $\mathbf{S_2}$ or $\mathbf{S_3}$. Compare Figs.~\ref{2nd_3rd_nm}-(b) with~\ref{3mark}-(b) and (c). 
An important feature to be noted here is that, like the situations of $\mathcal{A}_3^{S_i}$ with two of the environments being non-Markovian,
here also (i.e., for $\mathcal{A}_4^{S_i}$), the steady state is not attainable within a significantly large duration of time. 

\subsection{Remarks}

The results of this section lead us to 
conclude 
that replacement of any two Markovian environments including the one attached to the cold qubit is beneficial, as such an action  can enhance  transient cooling as compared to the corresponding ideal setups.
On the other hand, replacing  any one or both the  Markovian environments attached to the second and third qubits by  non-Markovian 
finite-size spin-environment(s), while keeping the cold qubit attached with a Markovian environment, results in a deterioration of performance  with respect to the ideal absorption refrigerator (i.e., the one with three Markovian environments). In other words, 
the investigations of all the situations $\mathcal{A}_j^{S_i}$ where $j=1$ to $4$ and $i=1$ to $3$, tell us  that one can attain a sufficient advantage in cooling of qubit $1$ (the cold qubit) whenever it is connected to a non-Markovian environment, whatever be the nature of the environments attached to the other two, with the advantage being over the ideal case of all three qubits being connected to Markovian environments. 
We can call these beneficial altered setups of the quantum cooling process as ``transient effects in cooling due to finite environments''. 
The reduction of temperature of the cold qubit with this non-Markovian setups owes its origin to the non-Markovianity incorporated in the environment attached to the first qubit. This leads us to define a quantifier of non-Markovianity for a quantum absorption refrigerator, as discussed in the succeeding section.
\section{Witnessing non-Markovianity in quantum refrigerators}
\begin{figure*}
\centering
\includegraphics[width=8.5cm]{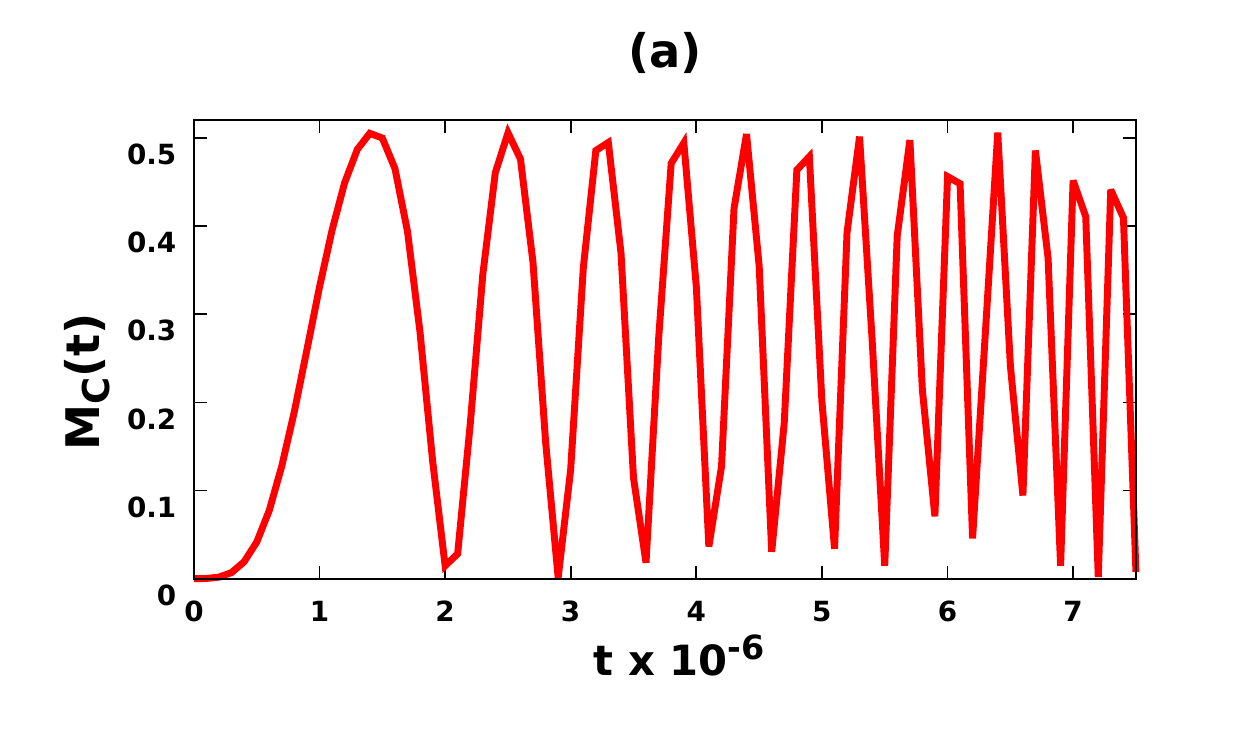}
\includegraphics[width=8.5cm]{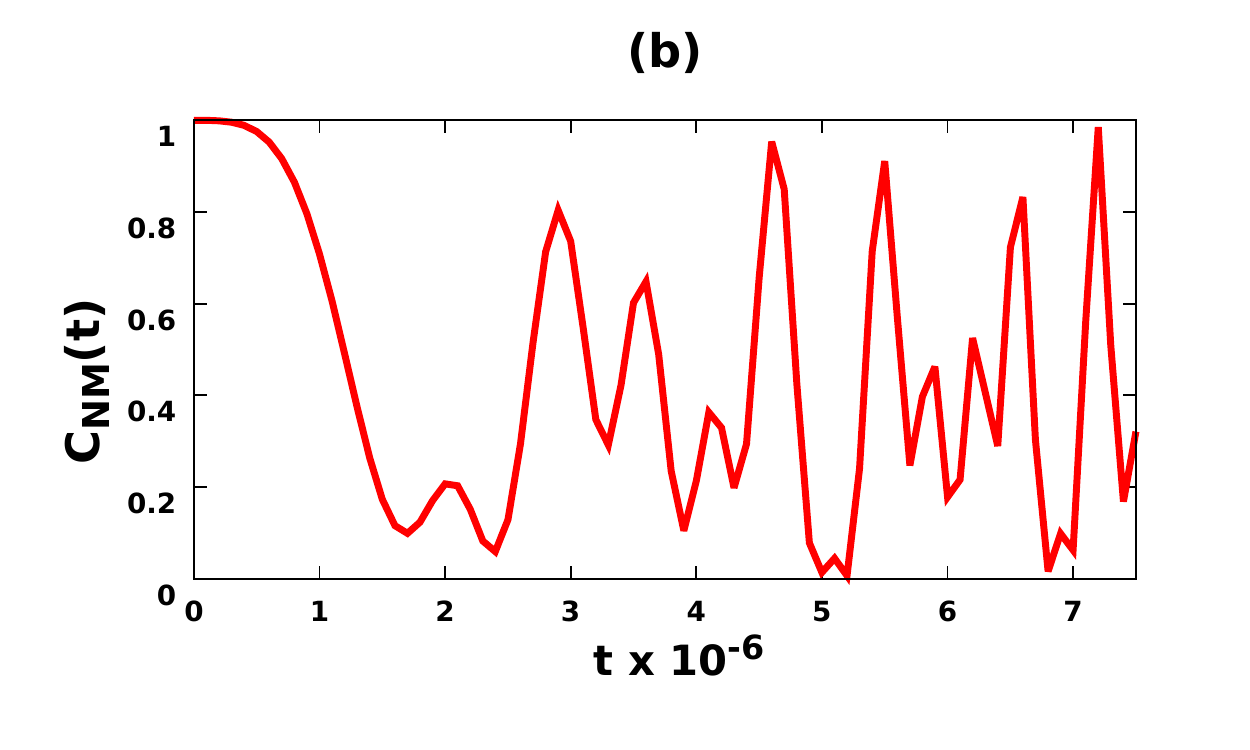}
\caption{Witness of non-Markovianity for a single-qubit single-bath refrigerator setup. Here in panel (a) the nature of the non-Markovianity witness $M_C(t)$ is plotted with time. In panel (b) we have depicted the system-auxiliary concurrence $(C_{NM}(t))$ with time, which shows a non-monotonic behavior. The quantity $M_C(t)$ is dimensionless, while the concurrence $(C_{NM}(t))$ is in ebits. The quantities plotted along the horizontal axes are dimensionless.
}
\label{nm_quantify_a1}
\end{figure*}
\label{Sec:4}
The results in the preceding sections have already confirmed that the efficiency of a quantum absorption refrigerator can be amplified by attaching the  cold qubit to a non-Markovian spin-environment, instead of a bosonic Markovian one. 
In this section, we further investigate the origin of this enhancement of efficiency.
For studying the cooling capacity of a three-qubit three-bath quantum refrigerator, we concentrate on the local temperature of qubit $1$ (cold qubit). Hence, to examine the origin of the reduction of cold qubit temperature in the altered beneficial situations discussed above, as compared to the corresponding ideal scenarios, we will only focus on the time dynamics of the first qubit by tracing out the other qubits (and all the environments).

Let us suppose that the single qubit system, named as the cold qubit, is described by the Hamiltonian $H_{S_1}^{\prime}=KE_1^{+}\ketbra{0}{0}+KE_1^{-}\ketbra{1}{1}$, and that the initial state of the system is $\rho^{1}_0$, which is diagonal in the eigenbasis of 
$H_{S_1}^{\prime}$. A non-Markovian channel, $N^t_{NM}$, is now applied on the system for time $t$, which results an evolution of the system such that the density matrix of the state remains diagonal in the eigenbasis of $H_{S_1}^{\prime}$, at each time $t$. Let the corresponding density matrix be given by $\rho^{1}_{NM}(t)=N^t_{NM}(\rho^{1}_0)=p_{NM}(t)\ketbra{0}{0}+(1-p_{NM}(t))\ketbra{1}{1}$. Similarly, if the system evolves through a Markovian channel $N^t_{M}$, one can obtain the dynamical state as $\rho^{1}_{M}(t)=N^t_{M}(\rho^{1}_0)=p_M(t)\ketbra{0}{0}+(1-p_M(t))\ketbra{1}{1}$. Hence, the temperatures, at time $t$, corresponding to the non-Markovian $(T_{NM}^t)$ and Markovian $(T_{M}^t)$ processes are defined through the relations,
\begin{eqnarray}
    &&\bra{0}\rho^{1}_{NM}(t)\ket{0} = \frac{e^{-\frac{E^{+}_1}{ T_{NM}^t}}}{e^{-\frac{E^{+}_1}{ T_{NM}^t}}+e^{-\frac{E^{-}_1}{T_{NM}^t}}}= \lambda_{NM}  \nonumber  \\
    \text{and}\quad &&\bra{0}\rho^{1}_{M}(t)\ket{0} = \frac{e^{-\frac{E^{+}_1}{ T_{M}^t}}}{e^{-\frac{E^{+}_1}{T_{M}^t}}+e^{-\frac{E^{-}_1}{T_{M}^t}}}= \lambda_M,
    \label{eq:lambda}
\end{eqnarray}
respectively.
Inverting these equations, one arrives at the difference of the dynamical temperatures of the system, when acted upon by a Markovian and a non-Markovian channel:   
\begin{equation}
\label{nm_measure}
 T_{M}^t-T_{NM}^t=(E^{+}_1-E^{-}_1)\Big[ \frac{1}{\ln\big(\frac{1}{\lambda_{M}} -1\big)} - \frac{1}{\ln\big(\frac{1}{\lambda_{NM}} -1\big)} \Big].
\end{equation}
This difference, $T_{M}^t-T_{NM}^t$, is a function of the parameters $\alpha_1, \alpha_2$ and $\alpha_3$, since a Markovian channel, \(N_M^t\), is an implicit function of these parameters. 
So, for a particular non-Markovian channel, this difference of temperatures can witness 
the deviation from Markovianity when 
the temperature $T_{M}^t$ is chosen to be the minimum temperature obtained among all  Markovian channels. 
Therefore, to obtain 
 a non-Markovianity witness, $M_C(t)$, for an arbitrary channel $C$, we optimize over all the Markovian channels in Eq.~(\ref{nm_measure}), which leads to the optimization over the channel parameters, $\alpha_1, \alpha_2$, and $\alpha_3$. The quantity $M_C(t)$ is defined as 
\begin{eqnarray}
    &&M_C(t)=\max\big\{0,\text{min}_{M}(T_{M}^t-T_{NM}^t)\big\}\nonumber\\
    &&=\max\Big\{0,(E^{+}_1-E^{-}_1)\Big[\frac{1}{\ln\big(\frac{1}{\lambda^{\prime}_{M}} -1\big)} - \frac{1}{\ln\big(\frac{1}{\lambda_C} -1\big)} \Big]\Big\}.\nonumber\\
\end{eqnarray}
Here the minimization is obtained over all the Markovian channels and $\lambda_M^{\prime}$ is obtained from Eq.~(\ref{eq:lambda}) for the minimum $T_{M}^t$.
This witness $M_C(t)$ may return a positive value when the channel is non-Markovian.
The  definition of the non-Markovianity witness involves an optimization over all possible Markovian channels. However, in the particular case that we consider, the system evolves using the Lindblad master equation, which acts as the Markovian channel. So here, the only relevant channel parameter to optimize over is the interaction strength between the system qubits and the Bosonic environments. This optimization does not cover the full set of Markovian channels, rather cover a subset of it. Hence, there is a possibility that a non-zero value of the witness would suggest that some Markovian channel may exist that provides the same cooling. 
An important point to be noted is that $M_C(t)$ is not a necessary and sufficient witness of non-Markovianity. It only provides a sufficient condition, which states that if $M_C(t)>0$, at any time $t$, the evolution is non-Markovian. However, if $M_C(t)=0$, it does not definitively determine whether the evolution is Markovian or non-Markovian. The nature of the quantity $M_C(t)$ with time for a single-qubit system evolving under a non-Markovian spin environment, discussed in Sec.~\ref{1q}, is demonstrated in Fig.~\ref{nm_quantify_a1}-(a). It is evident from the figure that this witness yields a positive value for $t>0$, indicating non-Markovianity. Therefore, the newly introduced witness, $M_C(t)$, effectively characterizes the non-Markovian nature inherent in the system.

It is  well-established  that  witnesses of non-Markovianity proposed in the literature are not all equivalent. However, it may still be worthwhile to consider a different witness of non-Markovianity, viz. the RHP witness~\cite{RHP} (see also~\cite{RHP0,BLP,Chrusci,Zheng,Debarba,Strasberg,Das_Roy,Huang_Guo} for non-Markovianity measures), and compare its dynamics with that of 
the 
witness, $M_C(t)$, presented here. 
The RHP witness of non-Markovianity is based on the monotonic decay of system-auxiliary entanglement under the Markovian dynamics of the system. If the system-auxiliary entanglement exhibits a non-monotonic pattern over time, it indicates that the dynamics of the system possesses non-Markovian behavior.  

We take a two-qubit system-auxiliary maximally entangled state of the form $\ket{\phi^+}=\frac{1}{\sqrt{2}}(\ket{00}+\ket{11})$, where the first qubit belongs to the system, and the second qubit belongs to the auxiliary. We now couple a non-Markovian spin environment, described in Eq.~(\ref{nm_bath}) for $N=2$, with the system qubit and evolve the system-auxiliary composite setup with time. The interaction between system and the environment is the same as in Eq.~(\ref{nm_int}). The initial entanglement between the two qubits is maximal, where we use the quantity ``concurence'' to measure the bipartite entanglement~\cite{Hill_Wootters,Wootters}. If we now observe the time-dynamics of the system-auxiliary concurrence, $(C_{NM}(t))$, depicted in Fig.~\ref{nm_quantify_a1}-(b), we can see that there is a distinct  non-monotonic behavior of entanglement, providing evidence of non-Markovianity. 
Hence, the proposed witness, $M_C(t)$, at least in certain scenarios, aligns with the RHP witness, highlighting its efficacy in capturing non-Markovian behavior. 

\section{Impact of Markovian noise on non-Markovian quantum refrigerators}
\label{Sec:5}
The study of non-Markovian quantum absorption refrigerators may sometimes be more realistic 
than the ideal refrigerator setup, as the environments are often likely to be non-Markovian and some such non-Markovian environments may not have any Markovian limit. Moreover, we realized in the preceding sections that incorporation of non-Markovian environments, whether by choice or necessity, can be advantageous. Noise or fluctuations is ubiquitous in physical systems, and can have non-negligible effects on the efficiency of devices. We now intend to study a situation  where the qubits of a non-Markovian quantum  refrigerator are affected by noise, and where we wish to investigate the effects of the same on its cooling capacity.
We consider two models of 
noise for analyzing the decay in the cooling capacity.\\
\\

\begin{figure*}
\includegraphics[width=8cm]{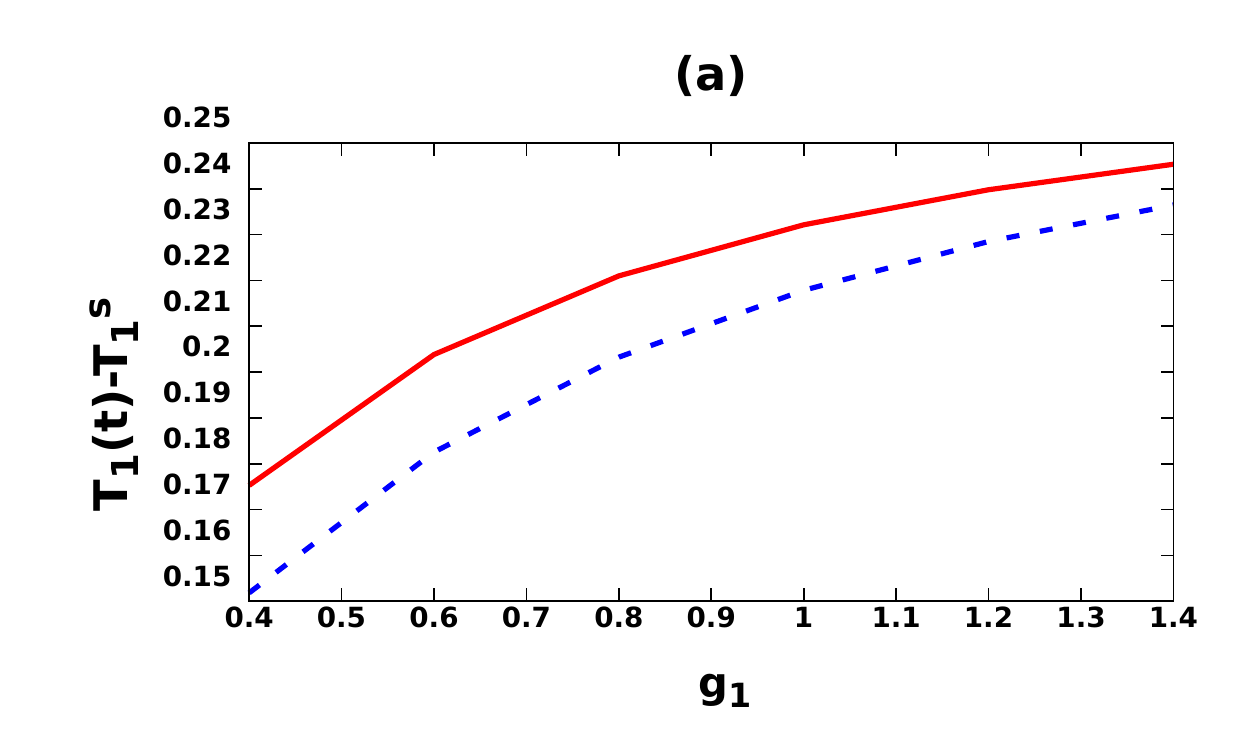}
\includegraphics[width=8cm]{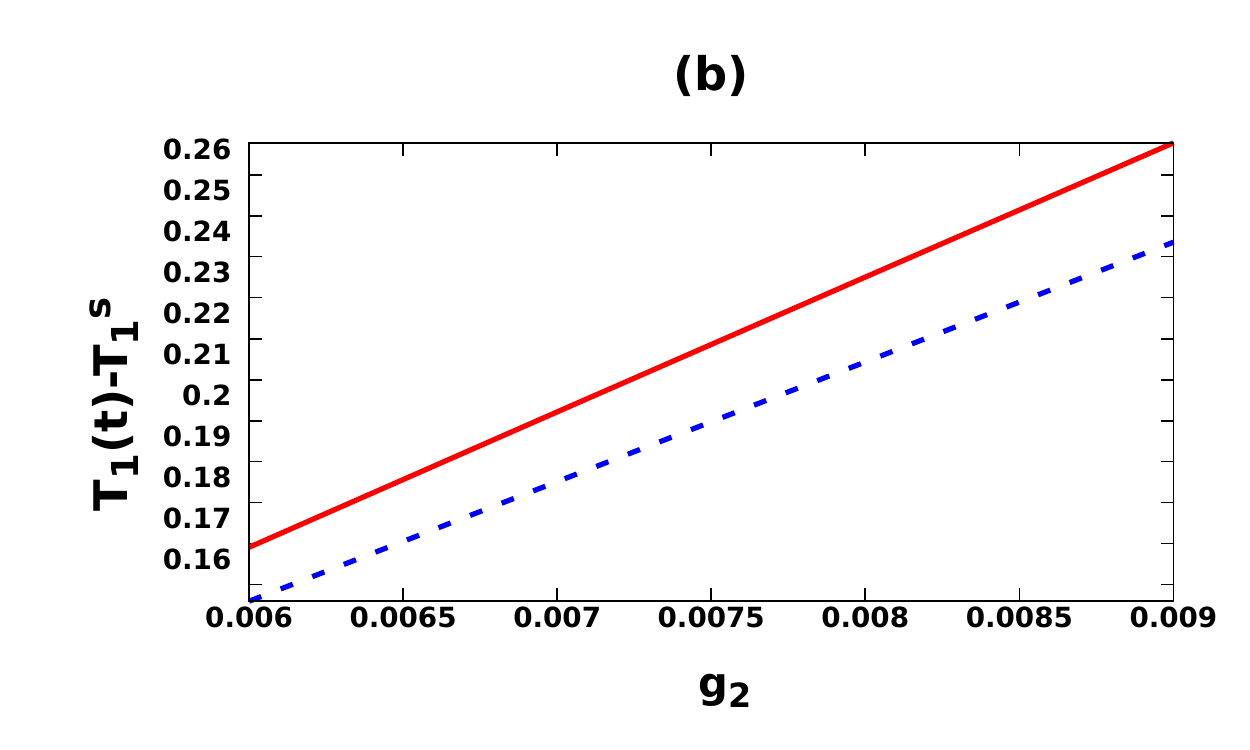}
\caption{
Effects of different types of Markovian noise on non-Markovian refrigerators. Here we depict the difference in the cold qubit temperature ($T_1(t)$) for the noisy setups (a) $\mathcal{N}_1(\mathcal{A}_1^{S_3})$ and (b) $\mathcal{N}_2(\mathcal{A}_1^{S_3})$ from the steady-state temperature ($T_1^s$) obtained in the noiseless $\mathcal{A}_1^{S_3}$ setup with the noise strength $g_1$ and $g_2$ respectively. The red solid curve gives the variation of the temperature difference at time $t=t_s=0.01$ in panel (a) and at time $t=t^{\prime}_s=0.05$ in panel (b). The blue dashed curve represents the same difference at time $t_s/2$ and $t_s^{\prime}/2$ in panels (a) and (b) respectively. The parameters chosen are $\tau_{N_1}=\tau_{N_2}=1.0$, $\alpha_{N_1}=0.001$, and 
$\alpha_{N_2}=0.01$. 
The quantities plotted along both the axes are dimensionless.}
\label{noise_scale}
\end{figure*} 
 




\subsection{Noise model I: Amplitude damping}
\label{N1}
Let us first consider the 
altered situations which provide better efficiencies than the ideal quantum absorption refrigerators with three qubits and three Markovian baths. The previously discussed altered situations, $\mathcal{A}_1^{S_i}$ and $\mathcal{A}_3^{S_i}$ for $i=1$, $2$, and $3$, are beneficial in comparison to the corresponding ideal setups, and can be termed as non-Markovian quantum refrigerators. In all these scenarios, the cold qubits are coupled to non-Markovian environments. 
We now suppose that along with being connected to a non-Markovian environment, the cold qubit also undergoes a Markovian quantum channel 
$\mathcal{N}_1(\cdot)$, where the argument is the state of the cold qubit in an altered situation 
(a non-Markovian refrigerator with a higher cooling capacity than the corresponding ideal scenario). In this paper, we have analyzed the noisy scenario 
corresponding to \(\mathcal{A}_1^{S_3}\), and refer to it as $\mathcal{N}_1(\mathcal{A}_1^{S_3})$. 
The cold qubit is therefore coupled to a non-Markovian spin-environment and also to a Markovian bosonic environment $B^{M}_{N_1}$ described by Eq.~(\ref{eq:Mar_bath}) with $i=N_1$. The other two qubits are connected to Markovian bosonic baths as before.
Therefore, the configuration is $\big\{B^{M}_{N_1} B_1^{NM},B_2^M,B_3^M\big\}$,
and the interaction Hamiltonian of the cold qubit and the Markovian environment $B^{M}_{N_1}$ is given by 
\begin{equation}
\label{N1_int}
    H_{SB^{M}_{N_1}}=\sqrt{g_1}\int_0^{\Omega} \hbar \sqrt{\tilde{\omega}}  d\omega  h_{N_1}(\omega)(\sigma^-_1 \eta_{\omega}^{N_1\dagger}+\sigma^+_1\eta_{\omega}^{N_1}).
\end{equation}
Here $g_1$ is the dimensionless noise strength. All other quantities are the same as in Eq.~(\ref{mark_int}), with $i$ replaced by $N_1$ for the bath parameters, and the $1$ in the subscript of $\sigma^{+}$ and $\sigma^{-}$ matrices refers to the $1^{\text{st}}$ qubit.
The dynamical equation of the density matrix of the three-qubit composite system for this noisy scenario
is the same as in Eq.~(\ref{QME_mixed}), with an extra dissipative term coming from the contribution of the noisy Markovian environment. So, the GKSL master equation for this noisy scenario turns out to be
\begin{equation}
\label{noise1}
    \frac{\partial \tilde{\rho}_s^{\mathcal{N}_1}(t)}{\partial t}=\mathcal{L}^{\prime}(\tilde{\rho}_s^{\mathcal{N}_1}(t))
    + \frac{\hbar}{K}g_1\mathcal{D}_{N_1}(\tilde{\rho}_s^{\mathcal{N}_1}(t)),
\end{equation}
for $i_m=2$, $3$ and $i_n=1$. Here $\tilde{\rho}_s^{\mathcal{N}_1}(t)$ is the density matrix of the three-qubit system
at time $t$. The dissipative term $\mathcal{D}_{N_1}(\tilde{\rho}_s^{\mathcal{N}_1}(t))$ is represented by Eq.~(\ref{Lindblad}), with $i=N_1$ and the state $\tilde{\rho}_s(t)$ being replaced by $\tilde{\rho}_s^{\mathcal{N}_1}(t)$. The jump operators and the transition rates associated with the term $\mathcal{L}^{\prime}(\tilde{\rho}_s^{\mathcal{N}_1}(t))$ for $i_m=2$ and $3$ are given in Appendix~\ref{appen:1}. The jump operators and the transition rates for the additional dissipative term $\mathcal{D}_{N_1}(\tilde{\rho}_s^{\mathcal{N}_1}(t))$, will be $L_{N_1}^{\omega^{\prime}}=L_1^{\omega^{\prime}}$ and $\{\gamma_{N_1}(\omega^\prime)\}=\{\gamma_{1}(\omega^\prime)\}$. These are also presented in Appendix~\ref{appen:1}. The temperature of the Markovian bath corresponding to amplitude damping noise is taken to be $1.0$ in units of $K/k_B$. This is chosen such that the initial state of the cold qubit is in equilibrium with the noisy Markovian bath and the spin environment at the same temperature.

In the noisy scenario, $\mathcal{N}_1(\mathcal{A}_1^{S_3})$, we see that the system attains a steady state, and the temperature of the cold qubit in the steady state is a function of the coupling strength $g_1$.  
When the parameter $g_1$ is reduced to a very small value, approximately $10^{-4}$, the steady-state temperature of the system $\mathcal{N}_1(\mathcal{A}_1^{S_3})$ closely approaches that of the noiseless system $\mathcal{A}_1^{S_3}$, which is approximately $0.7$, as illustrated in  Fig.~\ref{1st_nm}-(b). As the parameter $g_1$ increases, the resulting steady-state temperature of the cold qubit exceeds $0.7$, and for any value of $g_1>10^{-4}$, the temperature of the cold qubit does not fall below $0.7$.
So, the presence of the Markovian noise deteriorates the operation of the refrigerator when compared to the noiseless scenarios, although it still maintains its functionality as a refrigerator within a limited range of values for the parameter $g_1$. 
If a single qubit system interacts with a thermal environment, having the temperature $\tau_{N_1}=0$, via the interaction given in Eq.~(\ref{N1_int}), then the time-dynamics of the reduced system can be represented by the well-known amplitude damping channel.

In Fig.~\ref{noise_scale}-(a), we depict the  difference between the cold qubit temperature $(T_1(t))$ for the noisy scenario $\mathcal{N}_1(\mathcal{A}_1^{S_3})$ and the steady-state temperature obtained in the noiseless case $\mathcal{A}_1^{S_3}$ ($T_1^s\approx0.7$)
vs. the noise strength $g_1$. The solid red curve is for  time $t=t_s=0.01$, which corresponds to the time at which  the system reaches the steady-state regime. The blue dashed line represents the same at $t=\frac{t_s}{2}$. We can see that there is a finite gap between the two curves for small $g_1$, and this difference gradually diminishes as $g_1$ increases until it reaches a critical point at $g_1\approx 1.4$. Beyond this particular value of $g_1$, the system ceases to function as a refrigerator.

\subsection{Noise model II: Depolarizing}
\label{N2}
We  perform the same analysis with a different noisy channel $\mathcal{N}_2(\cdot)$. 
Here we again take the altered situation $\mathcal{A}_1^{S_3}$ and
the first qubit is kept in contact with a Markovian bosonic reservoir $B^{M}_{N_2}$, described in Eq.~(\ref{eq:Mar_bath}), in addition to the non-Markovian environment. Therefore, the configuration is now $\big\{B_{N_2}^MB_1^{NM},B_2^M,B_3^M\big\}$, and the interaction between the system and the noisy bath $B_{N_2}^M$ is given by
\begin{equation}
H_{SB^M_{N_2}} = \sqrt{g_2}\int_0^{\Omega}\hbar \sqrt{\tilde{\omega}} d\omega h_{N_2}(\omega) \sum_{X=x,y,z}\sigma_1^X \otimes (\eta_{\omega}^{N_2\dagger}+\eta_{\omega}^{N_2}). 
\end{equation}
Here, 
all the quantities have the same meanings as in Eq.~(\ref{mark_int}), and $g_2$ is a dimensionless coupling strength. The equation of motion of the density matrix of the three-qubit system for this situation is described by
\begin{equation}
    \frac{\partial \tilde{\rho}_s^{\mathcal{N}_2}(t)}{\partial t}=\mathcal{L}^{\prime}(\tilde{\rho}_s^{\mathcal{N}_2}(t))
    + \frac{\hbar}{K}g_2\sum_{X=x,y,z}\mathcal{D}_{N_{2_X}}(\tilde{\rho}_s^{\mathcal{N}_2}(t)).
\end{equation}
The explicit expressions of the jump operators and the transition rates of the dissipative terms $\mathcal{D}_{N_{2_X}}(\tilde{\rho}_s^{\mathcal{N}_2}(t))$ are given in Appendix~\ref{appen:2}. The temperature of the Markovian bath corresponding to depolarizing noise is considered to be $1.0$ in units of $K/k_B$. This is chosen such that the initial state of the cold qubit is in equilibrium with the noisy Markovian bath and the spin environment at the same temperature.
In this situation also, the noise model does not improve the performance of the refrigerator when compared to the noiseless scenario.
The temperature difference between those of the cold qubit ($T_1(t)$) in presence of the noisy channel, $\mathcal{N}_2$, and of the steady state  ($T_1^s$) obtained in the noiseless scenario 
is demonstrated against the dimensionless noise strength $g_2$ in Fig.~\ref{noise_scale}-(b).
The solid red curve is for the case when 
$t=t^{\prime}_s=0.05$,
whereas the blue dashed curve corresponds to the time $t=\frac{t^{\prime}_s}{2}$. \(t^{\prime}_s\) has the same meaning as \(t_s\) in the previous case with noise modelled by \(\mathcal{N}_1\). Just like the prior case of $\mathcal{N}_1$, this graph displays a pattern where the difference between the red and blue curves progressively increases in magnitude as the parameter $g_2$ rises until it surpasses the value of $0.009$, at which point the system no longer functions as a refrigerator.
From this noisy channel $\mathcal{N}_2$, the time dynamics of the well-known depolarizing channel can be restored if a single qubit system interacts with a thermal environment of temperature $\tau_{N_2} \rightarrow \infty$, via the interaction $H_{SB_{N_2}^M}$.

\section{Conclusion}
\label{Sec:6}
In this paper, we inspected a quantum refrigerator comprising of a few qubits, each of which are connected separately to local environments. We looked at situations where, in general, both TC and SSC coexist in  a quantum refrigerator system. We identified three domains, viz. where there is only TC, SSC better than TC, and TC better than SSC, considering different parameter regimes.

Along with
three-qubit quantum absorption refrigerators we also considered, for comparison and completeness,  single- and two-qubit self-contained thermal devices in presence of one or more spin-environments. In the single- and two-qubit cases, we note that although cooling is obtained under certain conditions, the oscillations in the final temperature are persistent throughout, and the envelopes never converge to a steady state value. Whereas, the three-qubit refrigerator attains a steady state when at most one of the bosonic baths is replaced by a spin environment.

We investigated the three-qubit absorption refrigerator for the three domains mentioned above, in presence of one or more spin environments. We studied the system when the environment of a single qubit, two qubits or all the environments of the three qubits of the refrigerator are replaced by non-Markovian spin environments.
The core objective of this paper was to show that replacing the Markovian bath attached to the cold qubit of a refrigerator with a non-Markovian reservoir, results in a considerable lowering of the temperature of the cold qubit, compared to the situation when all the three baths are Markovian. This advantage is apparent both in the transient and steady states. 
Note that in the steady state, the lower temperature of the cold qubit is obtained due to the backflow of information due to the finite size of the cold bath.

Connecting the other two qubits with non-Markovian environments, while keeping the cold qubit in a Markovian reservoir, does not provide advantage over the scenario where all the three baths are Markovian. Our focus is on the impact of backflow in finite-size environments, and our findings show that this backflow can enhance transient cooling in quantum refrigerators.
 
We have also suggested a way to gauge whether a channel is non-Markovian,
by defining a witness
of non-Markovianity, related to the non-Markovian refrigeration process. 
The refrigerator model has also been examined in the presence of two types of Markovian noise in addition to the Markovian and non-Markovian environments.
We identified the ranges of noise strength for which refrigeration still occurs.

 \acknowledgements 
 We acknowledge computations performed using Armadillo~\cite{Sanderson,Sanderson1},
   and QIClib~\cite{QIClib}
   on the cluster computing facility of the Harish-Chandra Research Institute, India. This research was supported in part by the `INFOSYS scholarship for senior students'. AG acknowledges support from the Alexander von Humboldt Foundation. We also acknowledge partial support from the Department of Science and Technology, Government of India through the QuEST grant (grant number DST/ICPS/QUST/Theme-3/2019/120).
   
\appendix
\section{Linblad operators and decay rates for the three-qubit qunatum absorption refrigerator}
\label{appen:1}
The Lindblad operators, corresponding to the GKSL equation presented in Eq.~(\ref{QME}) with the dissipative term as specified in Eq.~(\ref{Lindblad}), are expressed as 
\begin{eqnarray}
\label{eq:A1}
L^{E_1}_1 &=& \ketbra{111}{011}+\ketbra{100}{000} \nonumber \\
L^{E_1+g}_1 &=& \frac{1}{\sqrt{2}}\big(\ketbra{110}{+}+\ketbra{-}{001} \big)   \nonumber \\
L^{E_1-g}_1 &=& \frac{1}{\sqrt{2}}\big(\ketbra{+}{001}-\ketbra{110}{-} \big)   \nonumber \\
L^{E_2}_2 &=& \ketbra{110}{100}+\ketbra{011}{001}    \nonumber \\
L^{E_2+g}_2 &=& \frac{1}{\sqrt{2}}\big(\ketbra{111}{+}-\ketbra{-}{000} \big)   \nonumber \\
L^{E_2-g}_2 &=& \frac{1}{\sqrt{2}}\big(\ketbra{+}{000}+\ketbra{111}{-} \big)   \nonumber \\
L^{E_3}_3 &=& \ketbra{111}{110}+\ketbra{001}{000}    \nonumber \\
L^{E_3+g}_3 &=& \frac{1}{\sqrt{2}}\big(\ketbra{011}{+}+\ketbra{-}{100} \big)   \nonumber \\
L^{E_3-g}_3 &=& \frac{1}{\sqrt{2}}\big(\ketbra{+}{100}-\ketbra{011}{-} \big)   
\end{eqnarray}
Here $\ket{+}=\frac{1}{\sqrt{2}}(\ket{101}+\ket{010})$ and $\ket{-}=\frac{1}{\sqrt{2}}(\ket{101}-\ket{010})$.
The remaining nine Lindblad operators, representing the opposite processes, can be determined using the relation $ L^{-\omega^{\prime}}_i= L^{\omega^{\prime} \dagger}_i$. In the master equation associated with the Markovian qubit-bath interaction, the information about the reservoirs is encapsulated in the transition rates or decay constants, $\{ \gamma_{i}(\omega^{\prime}) \}$, which is given by
\begin{eqnarray}
\label{gamma}
\gamma_{i}(\omega^{\prime}) &=& J_i(\omega^{\prime}) [1+f(\omega^{\prime},\beta_i)] \quad \quad  \omega^{\prime}>0 \nonumber \\
                  &=& J_i(|\omega^{\prime}|) f(|\omega^{\prime}|,\beta_i) \quad \quad \quad   \omega^{\prime}<0,
\end{eqnarray}
where $f(\omega^{\prime},\beta_i)=[\exp(\beta_i \hbar \omega^{\prime})-1]^{-1}$ is the Bose-Einstein distribution function for the bosonic heat baths.


\section{Lindblad operators and the transition rates for the noise model II}
 \label{appen:2}

The Lindblad operators for the noise model II, with the bath configuration $\big\{B^M_{N_2}B_1^{NM},B_2^M,B_3^M\big\}$ and the interaction of the system with the noisy bath, given by $H_{SB_{N_2}^M}$, take the following forms
\begin{eqnarray}
 &&L_{N_{2_x}}^{E_1} = L^{E_1}_1, 
\quad L_{N_{2_x}}^{E_1+g} = L_1^{E_1+g}, 
\quad L_{N_{2_x}}^{E_1-g} = L_1^{E_1-g}, 
\nonumber \\
&& L_{N_{2_y}}^{E_1} = i(\ketbra{111}{011} + \ketbra{100}{000}), \nonumber \\
&& L_{N_{2_y}}^{E_1+g} = \frac{i}{\sqrt{2}}(\ketbra{-}{001}+\ketbra{110}{+}), \nonumber \\
&& L_{N_{2_y}}^{E_1-g} = \frac{i}{\sqrt{2}}(\ketbra{+}{001}-\ketbra{110}{-}), \nonumber \\
&& L_{N_{2_z}}^0 = \ketbra{000}{000}+\ketbra{001}{001}
+\ketbra{011}{011}  \nonumber \\
 &&\phantom{jini sokol} -\ketbra{100}{100}-\ketbra{110}{110}-\ketbra{111}{111}, \nonumber \\
 &&L^{2g}_{N_{2_z}} = -\ketbra{-}{+}, \quad \text{and} \quad 
 L^{-2g}_{N_{2_z}} = -\ketbra{+}{-}.
\end{eqnarray}
Also, $L_{N_{2_X}}^{-\omega}=L_{N_{2_X}}^{\omega^{\dagger}}$ gives the corresponding operators for the opposite processes. The transition rates $\{\gamma_{N_{2_X}}(\omega^{\prime})\}$ are given in Eq.~(\ref{gamma}) with $i=N_{2_X}$.

\end{document}